\documentclass{aa}

\usepackage{txfonts}
\usepackage{graphicx}
\usepackage{epsfig}

\begin{document}

\title{Radial mixing in protoplanetary accretion disks}
\subtitle{VII. $2$-dimensional transport of tracers}

\author{M. Wehrstedt \and H.-P. Gail}

\institute{Institut f\"ur Theoretische Astrophysik, Universit\"at Heidelberg, 
  Albert-\"Uberle-Str. 2, 69120 Heidelberg, Germany 
  (gail@ita.uni-heidelberg.de)}

\offprints{H.-P. Gail}

\date{Received XXX/ Accepted XXX}

\abstract{}
{The detection of significant concentrations of crystalline silicates in comets 
indicates an extensive radial mixing in the primordial solar nebula, i.e. the 
protoplanetary accretion disk of our solar system. In studying the radial transport
of matter within protoplanetary disks by numerical model calculations it is
essential to resolve the vertical disk structure since matter is mixed
radially inward and outward by a complex $2$-dimensional flow pattern within the disk
that is superposed on the global inward directed accretion flow. It is further
essential to follow numerically the advection-diffusion processes over a period
of at least 10$^6$ yrs to allow for a full development of the radial 
concentration profile built up by radial mixing from the warm inner to the
cool outer parts of protoplanetary disks beyond of 10 AU.}
{Numerical model calculations of protoplanetary accretion disks with
radial and vertical mixing are performed by solving a set of $2$-dimensional
transport-diffusion-reaction equations for some important tracers self-consistently
with the set of disk equations in the $1$+$1$-dimensional approximation. The
global $2$D velocity field of the disk is calculated from an approximate analytical
solution for the meridional flow pattern, which exhibits an inward drift in
the upper layers and an outward drift in the midplane in most parts of the disk.
The disk model is based on the $\beta$-prescription of viscosity and
considers vertical self-gravitation of the disk. This kind of semianalytical
approximations allows with presently available computer capacity to follow the
evolution of the disk and the transport and mixing of tracers in vertically
resolved 2D-models over the required long periods of disk evolution. The mixing
processes in the disk are studied for the following species: amorphous silicate
grains (forsterite, enstatite) which crystallise by annealing in the warm inner
parts of the disk, and carbonaceous grains which are destroyed by surface
reactions with ${\rm OH}$ molecules at elevated temperatures.
}
{Considerable fractions of crystallised silicates and methane (formed as a 
by-product of carbon combustion) are transported to the
site of comet formation far from the protosun within a period of 10$^6$ yrs.
The $2$-dimensional transport of tracers in the solar nebula therefore offers
a natural explanation for the presence of crystalline silicates in comets and
the significant portions of crystalline silicates observed in
accretion disks around young stellar objects.
}
{}

\keywords{Accretion, accretion disks -- solar system: formation -- dust, 
  extinction}

\maketitle


\section{Introduction}

In the standard one-zone model of protoplanetary accretion disks 
(Pringle~\cite{pri81}; Lin \& Papaloizou~\cite{lin85}) the vertically averaged 
flow field of the disk is characterized by a radial inward drift of the disk
matter within the  inner, chemically active zone of the disk. 
Only in the icy region far away from the proto-sun the disk's flow field is 
directed outward. However, Urpin~(\cite{urp84}) found in his analytical work
large-scale meridional flow patterns to exist. This meridional flow field is
characterized by an {\it outward} directed drift of the disk matter close to the
disk midplane whereas the flow is directed {\it inward} in higher layers. The
result of Urpin~(\cite{urp84}) has been confirmed by several
authors by different analytical, semi-analytical and numerical methods 
(Siemiginowska~\cite{sie88}; Kley \& Lin~\cite{kle92}; R\'o\.zyczka et 
al.~\cite{roz94}; Klu\'zniak \& Kita~\cite{klu00}; Regev \& 
Gitelman~\cite{reg02}; Tscharnuter \& Gail \cite{Tsa07}). Thus a meridional flow field seems to be the universal
type of flow pattern existing in protoplanetary accretion disks. 

With respect to radial mixing processes in protoplanetary disks the structure 
of the flow field is of utmost importance. This is a consequence of the fact
that the average radial transport of matter in protoplanetary disks by
advection occurs on a similar timescale as the transport by turbulent diffusion,
namely on the viscous timescale. Hence, besides a realistic description of the
turbulent diffusion, an exact knowledge of the large-scale flow field in disks
is essential for calculating the transport of matter within disks. Here we consider the effect of meridional flows. Another type of hydrodynamic mixing associated with gravitational instabilities is discussed by Boss (\cite{bos04,bos07,bos08})
and found to be very efficient in that case. 

Previous model calculations of protoplanetary disks which include the 
calculation of radial mixing of species have considered only the vertically
averaged one-zone velocity field as the flow field of the disk (Stevenson \&
Lunine~\cite{ste88}; Cyr et al.~\cite{cyr98}; Drouart~\cite{dro99};
Bockel\'ee-Morvan et al.~\cite{boc02} as well as the series of papers of the
ITA group: Gail~\cite{gai01}; Wehrstedt \& Gail~\cite{weh02};
Gail~\cite{gai02}; Gail~\cite{gai04}; Wehrstedt \& Gail~\cite{weh03}, henceforth
called Papers I -- V). The only exception are the $2$-dimensional model
calculation of Keller \& Gail~(\cite{kel04}, henceforth called Paper VI)
and Tscharnuter \& Gail (\cite{Tsa07}) where for the first time the meridional
velocity field is used for the computation of the radial mixing of species in
protoplanetary disks. The results show that the outward transport of species is
much more efficient in the meridional flow field than in the one-zone flow
field. 

In the present work we extend the work of Paper VI. This is done by improving
the time-dependent one-zone models of Papers II and V by calculating the disk
structure in the $1$+$1$-dimensional approximation (e.g. Lin \& 
Papaloizou~\cite{lin85}) simultaneously with the $2$-dimensional mixing of
species in the disk. We therefore accept the small deviations of the
$1$+$1$-dimensional disk structure from the disk structure of the exact
$2$-dimensional hydrodynamic calculations of Paper VI in order to save
computing time. In contrast, however, the present $1$+$1$-dimensional disk model
is coupled with a sophisticated chemical model which includes the calculation of
equilibrium condensation of the most abundant solids, annealing of silicates,
and combustion of solid carbon. Furthermore, the present disk model includes a
detailed opacity calculation considering the Rosseland and Planck opacity means
of the most important absorbers in the disk. 

With this powerful tool we investigate radial mixing processes in protoplanetary
disks. In particular the radial mixing of species in the solar nebula is of
great interest to explain the composition of the most pristine solar system 
bodies, i.e., the comets. From observations it is long known that some fraction
of the dust in comets is crystalline (e.g. Swamy et al. \cite{Swa88}; Hanner et
al. \cite{han97}) and crystalline silicate grains have been detected in
accretion disks around young stellar objects (e.g. Meeus et al. \cite{Meu01};
Bouwman et al. \cite{Bou01}; van Boekel et al. \cite{vBo04,vBo05}; Keller et al.
\cite{Kel05}). Both these observations are interpreted as resulting from mixing
material from the central parts of the disk to the outer regions though also
other processes have been invoked for that (see, e.g., Alexander et al. 
\cite{Ale07}; Wooden et al. \cite{woo07} for a discussion).

The paper is organized as follows: Section~\ref{sectra} presents the treatment
of the $2$-dimensional transport of tracers in the disk within the disk model. 
In Sects.~\ref{secrad} and \ref{secver} the method of calculating the radial and
vertical disk structure is described. Section~\ref{secvel} addresses the
calculation of the velocity field of the disk, in particular the meridional
flow field. Section~\ref{secnum} deals with the numerical treatment of the
disk model. In Sect.~\ref{secres} we present the results and finally make our
conclusions in Sect.~\ref{seccon}.

\section{Transport of tracers}\label{sectra}

\subsection{The transport-reaction equation}

The transport of tracers with small concentration embedded in a carrier medium
is given by the transport-reaction equation (Hirschfelder et al. \cite{hir64};
cf. Paper VI) 
\begin{equation}\label{dncidt}
\frac{\partial}{\partial t}\, nc_{i} + \vec{\nabla} \cdot nc_{i}\vec{v}_{i} = 
  \vec{\nabla} \cdot nD_{i}\, \vec{\nabla} c_{i} + R_{i} \ , 
\end{equation}
where $n$ denotes the particle density of the carrier medium, 
$\vec{v}_{i}$ the velocity vector of tracer component $i$, $D_{i}$ the binary 
diffusion coefficient of tracer $i$ relative to the carrier, $R_{i}$ the 
rate term of gains and losses by chemical reactions of tracer $i$, and 
\begin{equation}\label{ci}
c_{i} = \frac{n_{i}}{n}
\end{equation}
the concentration of tracer $i$. $n_{i}$ denotes the particle density of tracer
$i$. The second term on the l.h.s. of Eq.~(\ref{dncidt}) is the advection term. The first term on the r.h.s. the diffusion term. 

By using the continuity equation 
\begin{equation}\label{3dkontiglg}
\frac{\partial n}{\partial t} + \vec{\nabla} \cdot n\vec{v} = 0
\end{equation}
the transport-reaction equation~(\ref{dncidt}) can be changed to 
\begin{equation}\label{dcidtgen}
\frac{\partial c_{i}}{\partial t} + \vec{v} \cdot \vec{\nabla} c_{i} = 
  \frac{1}{n}\, \vec{\nabla} \cdot nD_{i}\, \vec{\nabla} c_{i} + 
  \frac{R_{i}}{n} \ . 
\end{equation}
For simplicity 
\begin{equation}\label{vi}
\vec{v}_{i} = \vec{v}
\end{equation}
is assumed, i.e., all tracer velocities $\vec{v}_{i}$ equal the velocity
$\vec{v}$ of the carrier medium. For the present model calculations this is an
acceptable assumption since the considered tracers are micron-sized dust grains
which are carried along with the flow in most parts of the disk. More precisely,
for micron-sized particles Eq.~(\ref{vi}) only holds in the densest and
chemically active disk regions whereas in the less dense outskirts of the disk
the motions of carrier gas and tracers may decouple, i.e., $\vec{v}_{i} \neq
\vec{v}$. 

The binary diffusion coefficient is assumed to be given by 
\begin{equation}\label{di}
D_{i} = \frac{\nu}{S_{i}} \ , 
\end{equation}
where $\nu$ is the kinematic viscosity of the disk matter and $S_{i}$ the 
Schmidt number of tracer $i$. The Schmidt number is set in the present calculations to  
\begin{equation}\label{si}
S_{i} = 1
\end{equation}
for all tracers. The approximation for the Schmidt number (\ref{si}) is based on
the same approximations as assumption~(\ref{vi}). A Schmidt number equal to unity
means that the tracer moves as the carrier gas does. In less dense parts of
the disk this approximation fails, and $S_{i} > 1$.

Since regions of low density do not contribute much to the total tracer
transport, we apply (\ref{vi}) and (\ref{si}) for all tracers in the present
model. The  value of $S_i$ can, however, not be fixed with any precision because
of the  unclear physics of the viscous transport in accretion disks. For some
discussions on the value of $S_i$ see, e.g., Johansen et al. 
\cite{joh05,joh06}; Turner et al. \cite{tur06}; Pavlyuchenkov \& 
Dullemond \cite{pav07}.

For calculating the tracer transport in two dimensions we assume axial symmetry 
and use polar coordinates ($r',\theta$). Here, $r'$ denotes the polar radius and 
$\theta$ the polar angle measured from the midplane of the disk. With this
choice of coordinates the transport-reaction  equation~(\ref{dcidtgen}) takes
the form 
\begin{eqnarray}\label{dcijdtpol}
\frac{\partial c_{i,j}}{\partial t} & \!\! + \! & v_{r}' 
  \frac{\partial c_{i,j}}{\partial r'} + v_{\theta} \frac{1}{r'}
  \frac{\partial c_{i,j}}{\partial \theta} = \frac{1}{r'^2 n}\, 
  \frac{\partial}{\partial r'}\, r'^2 n D\, 
  \frac{\partial c_{i,j}}{\partial r'} \nonumber \\
& \!\! + \!\! & \frac{1}{r'^2 n \cos{\theta}} 
  \frac{\partial}{\partial \theta} \cos{\theta}\, n D\, 
  \frac{\partial c_{i,j}}{\partial \theta} + \frac{R_{i,j}}{n} \ , 
\end{eqnarray}
where $v_{r}'$ and $v_{\theta}$ are the velocities in radial and polar 
direction, respectively. 
The index $j$ denotes a kind of tracer occurring in a special modification $i$ 
(see following section).

\subsection{The set of tracers}

The tracers which are considered in the model calculations are: 
\begin{enumerate}
\item silicate dust grains (forsterite and enstatite) of different degrees of 
  crystallisation and 
\item carbon dust grains of different sizes. 
\end{enumerate}
The tracers experience the following reactions: 
\begin{enumerate}
\item Silicate dust grains (forsterite and enstatite) start to anneal at 
  temperatures above $\sim 800\,{\rm K}$, i.e., the degree of crystallisation 
  increases until it reaches unity. 
\item Carbon dust grains become decomposed by reactions with ${\rm OH}$
molecules at the 
  grain's surfaces. This process critically depends on the density of ${\rm OH}$
  molecules in the gas phase and starts to operate at temperatures of
  $\sim 1\,100\,{\rm K}$ under conditions encountered in protoplanetary
  disks. 
\end{enumerate}
The rate terms for the basic reactions are given in Paper II. 

The set of transport-reaction equations~(\ref{dcijdtpol}) determines the 
concentration of crystalline forsterite, $c_{i,{\rm for}}$, and crystalline 
enstatite, $c_{i,{\rm ens}}$, with different degrees of crystallisation 
$x_{i}$, and the concentration $c_{i,{\rm car}}$ of solid carbon grains with
different grain sizes $a_{i}$.  The $c_{i,j}$ are used to calculate the average
degrees of crystallisation of forsterite and enstatite (cf. Paper II) 
\begin{equation}\label{fcryfor}
f_{\rm cry,for} = \sum_{i=1}^{I_{\rm sil}} x_{i}\, c_{i,{\rm for}} \Big/ 
  \sum_{i=1}^{I_{\rm sil}} c_{i,{\rm for}} \ , 
\end{equation}
\begin{equation}\label{fcryens}
f_{\rm cry,ens} = \sum_{i=1}^{I_{\rm sil}} x_{i}\, c_{i,{\rm ens}} \Big/ 
  \sum_{i=1}^{I_{\rm sil}} c_{i,{\rm ens}}\ ,
\end{equation}
and the degree of condensation of C into solid carbon 
\begin{equation}\label{fcar}
f_{\rm car} = \frac{1}{V_{0,{\rm car}}\epsilon_{\rm C}} 
  \sum_{i=1}^{I_{\rm car}-1} \frac{4\pi}{3} a_{i}^3 \,c_{i,{\rm car}} \ . 
\end{equation}
$V_{0,{\rm car}}$ denotes the volume of a carbon atom within solid carbon and
$\epsilon_{\rm C}$ the (solar) abundance of C. $I_{\rm sil}$ is the
number of sampling points for the discretised degrees of crystallisation
$x_{i}$, $i = 1 \dots I_{\rm sil}$, and $I_{\rm car}$ the number of sampling
points for the size spectrum of solid carbon grains $a_{i}$, $i = 1 \dots
I_{\rm car}$. The set of sampling points for $x_{i}$ and $a_{i}$ is chosen as
in Paper V ($I_{\rm sil} = 3$, $I_{\rm car} = 31$). 

For a more detailed discussion of the processes of silicate annealing and
carbon combustion the reader is referred to the other papers of this series, in
particular Paper I and II.

\subsection{Solution of the set of transport-reaction equations}

The set of transport-reaction equations~(\ref{dcijdtpol}) is discretised in 
first order in time and in second order in space with respect to the diffusion 
terms. The advection terms are treated by a standard upwind method (see Paper
II). 

The transport-reaction equations~(\ref{dcijdtpol}) are solved by an ADI 
({\bf A}lternating {\bf D}irection {\bf I}mplicit) method (Press et 
al.~\cite{pre92}). By ADI each transport-reaction equation is split into two
equations for separate directions in space: one equation
contains the terms in $r'$-direction and the other equation the terms in
$\theta$-direction. The rate term is split into two half-steps and equally
distributed on both directions. The resulting matrices of the separate equations
have tri-diagonal structure and thus can be inverted numerically fast and easily
(Press et al.~\cite{pre92}). Inverting the matrix of the unsplitted
transport-reaction equation~(\ref{dcijdtpol}), which has a band-tri-diagonal
shape, would be numerically much more time-consuming and therefore less
efficient than the ADI method. 

To ensure numerical stability the sequence of solution of the splitted 
equations is permuted between two successive time steps (therefore
{\it Alternating} Directions Implicit method). The transport-reaction equations
are solved fully implicit.

\section{Radial disk structure}\label{secrad}

The radial structure of the disk is calculated in the one-zone approximation by 
using the $\beta$-prescription of viscosity and by accounting for the 
vertical self-gravity of the disk. The resulting set of equations for the
radial disk structure is given by (cf. Paper V): 

\noindent
{\it 1.)} Time evolution of the surface density $\Sigma$: 
\begin{equation}\label{dsigmadt}
\frac{\partial\Sigma}{\partial t} = \frac{3}{r} \frac{\partial}{\partial r} 
  \sqrt{r} \frac{\partial}{\partial r} \nu\Sigma\sqrt{r} \ . 
\end{equation}
where $r$ is the radial distance from the mass-centre. 

\noindent
{\it 2.)} Keplerian angular velocity: 
\begin{equation}\label{omega}
\Omega = \frac{v_{\rm K}}{r}= \sqrt{\frac{GM_{\ast}}{r^3}} \ , 
\end{equation}
where $v_{\rm K}$ is the Keplerian velocity, $G$ the gravitational constant and 
$M_{\ast}$ the stellar mass. 

\noindent
{\it 3.)} $\beta$-viscosity: 
\begin{equation}\label{nubeta}
\nu_{\beta} = \beta\,r^2\Omega \ , 
\end{equation}
where $\beta$ denotes the viscosity parameter which is chosen to be $10^{-5}$ 
in the model calculations (cf. Paper V). 

\noindent
{\it 4.)} Isothermal sound speed: 
\begin{equation}\label{cs}
c_{\rm s} = \sqrt{\frac{k_{\rm B}T_{\rm c}}{\mu m_{\rm H}}} \ , 
\end{equation}
where $k_{\rm B}$ is the Boltzmann constant, $T_{\rm c}$ the temperature in the 
midplane of the disk, $\mu$ the mean molecular weight and $m_{\rm H}$ the 
proton mass, respectively. 

\noindent
{\it 5.)} Pressure scale height by accounting for the disk's vertical 
self-gravitation: 
\begin{equation}\label{hs}
h_{\rm s} = \frac{2\pi G\Sigma}{\Omega^2} \left[ \sqrt{ 1 + 
  \left( \frac{c_{\rm s}\Omega}{2\pi G\Sigma} \right)^2 } - 1 \right] \ . 
\end{equation}

\noindent
{\it 6.)} Mean vertically averaged mass density: 
\begin{equation}\label{rhom}
\rho_{\rm m} = \frac{\Sigma}{2h_{\rm s}} \ . 
\end{equation}

\noindent
{\it 7.)} Mean molecular weight: 
\begin{equation}\label{mu}
\mu = \frac{ \rho_{\rm m} k_{\rm B} T_{\rm c} }
  { m_{\rm H} ( p_{\rm H} + p_{\rm H_2} + p_{\rm He} ) } \,. 
\end{equation}
The calculation of the partial pressure $p_{\rm X}$ of species 
${\rm X}$ in chemical equilibrium is described in Paper I. 

\noindent
{\it 8.)} Rosseland and Planck means of the mass extinction coefficient: 
\begin{eqnarray}
\kappa_{\rm R/P,for}&=&f_{\rm cry,for}\kappa_{\rm R/P,for,cry}\nonumber
\\
&&\quad + \left(1 - f_{\rm cry,for} \right) \kappa_{\rm R/P,sil,am}
\label{kafor}
\\
\kappa_{\rm R/P,ens}&=&f_{\rm cry,ens}\kappa_{\rm R/P,ens,cry}\nonumber
\\
&&\quad+
 \left(1 - f_{\rm cry,ens} \right) \kappa_{\rm R/P,sil,am}
\label{kaens}
\\
\kappa_{\rm R/P,dust} & = & f_{\rm car}\kappa_{\rm R/P,car} + 
  f_{\rm ens}\kappa_{\rm R/P,ens} + f_{\rm for}\kappa_{\rm R/P,for} \nonumber \\
& &\quad + \ f_{\rm iro}\kappa_{\rm R/P,iro} + f_{\rm cor}\kappa_{\rm R/P,cor} 
\label{kadust}
\end{eqnarray}
\begin{eqnarray}
\kappa_{\rm R/P} &=& f_{\rm ice}\kappa_{\rm R/P,ice} + \left( 1 - f_{\rm ice} 
  \right) \kappa_{\rm R/P,dust} \nonumber
\\
&&\quad + \left( 1 - f_{\rm cor}\right)\kappa_{\rm mol} \ . 
\label{karp}
\end{eqnarray}
For the calculation of the opacity the main dust absorbers are considered in 
the disk model. The abbreviations used here denote: 'for` $\equiv$ (crystalline)
forsterite, 'ens` $\equiv$ (crystalline) enstatite, 'sil,am` $\equiv$ amorphous
silicate, 'car` $\equiv$ solid carbon, 'iro` $\equiv$ solid iron, 'cor` $\equiv$
corundum and 'ice` $\equiv$ water ice, 'mol` $\equiv$ molecules. $f_{\rm Z}$
denotes the degree of condensation of the key element of condensate ${\rm Z}$
(${\rm Si}$ for the silicates, ${\rm C}$ for solid carbon, ${\rm Fe}$ for solid
iron, ${\rm Al}$ for corundum and ${\rm O}$ for water ice). The method of
calculating $f_{\rm car}$, $f_{\rm iro}$ and $f_{\rm ice}$ is described in Paper
I and II, respectively, and of $f_{\rm for}$, $f_{\rm ens}$ and $f_{\rm cor}$ in
Paper V. $f_{\rm cry,for}$ and $f_{\rm cry,ens}$ are calculated by
Eqs.~(\ref{fcryfor}) and (\ref{fcryens}), respectively, and $f_{\rm car}$ by
Eq.~(\ref{fcar}). Analytical fit formulae for the Rosseland and Planck mean of
the opacity of the individual species are given in Wehrstedt~(\cite{weh03b}). 

\noindent
{\it 9.)} Rosseland and Planck mean vertical optical depth at the midplane: 
\begin{equation}\label{taurp}
\tau_{\rm R/P} = \frac{1}{2} \Sigma\,\kappa_{\rm R/P} \ . 
\end{equation}

\noindent
{\it 10.)} Viscous dissipation rate: 
\begin{equation}\label{dote}
{\dot E}_{\nu} = \frac{9}{8}\,\Omega^{2}\nu\Sigma \ . 
\end{equation}

\noindent
{\it 11.)} Effective temperature of the disk surface: 
\begin{equation}\label{teff}
\sigma T_{\rm eff}^{\,4} = \left( 1 + \frac{1}{4\tau_{\rm P}} \right) 
  {\dot E}_{\nu} + \sigma T_{\rm cloud}^{\,4} \ , 
\end{equation}
where $\sigma$ is the Stefan-Boltzmann constant and $T_{\rm cloud}$ the 
temperature of the ambient molecular cloud. 

\noindent
{\it 12.)} Temperature at the midplane: 
\begin{equation}\label{tc}
\sigma T_{\rm c}^{4} = \left( \frac{3}{4}\tau_{\rm R} + \frac{1}{4\tau_{\rm P}} 
  \right) {\dot E}_{\nu} + \sigma T_{\rm cloud}^{4} \ . 
\end{equation}

\noindent
{\it 13.)} Radial drift velocity: 
\begin{equation}\label{vr}
v_{r} = \frac{3}{\sqrt{r}\Sigma}\, \frac{\partial}{\partial r}\, 
  \nu\Sigma\sqrt{r} \ . 
\end{equation}

\noindent
{\it 14.)} Tracer transport in the one-zone approximation: 
\begin{equation}\label{dcijdt}
\frac{\partial c_{i,j}}{\partial t} + v_{r} \frac{\partial c_{i,j}}{\partial r} 
  = \frac{1}{rn} \frac{\partial}{\partial r}\, rnD\, 
  \frac{\partial c_{i,j}}{\partial r} + \frac{R_{i,j}}{n} \ . 
\end{equation}

\noindent
{\it 15.)} Rate of mass accretion: 
\begin{equation}\label{dotm}
{\dot M}(r,t) = 2\pi{r}\Sigma\,v_{r} \ . 
\end{equation}

\noindent
{\it 16.)} Variation of the stellar mass by accretion of matter onto the
protostar: 
\begin{equation}\label{mstar}
M_{\ast}(t) = M_{\ast}(0) - \int_{0}^{t} {\dot M}(r_{\rm in},t')\,dt' \ . 
\end{equation}
Note that $\dot M(r_{\rm in})$ is negative.

The set of equations for the radial disk structure (\ref{dsigmadt}) -- 
(\ref{mstar}) is solved with standard methods up to an accuracy of 
$\Delta_{\rm rad} = 10^{-5}$ for the radial model structure. 

\section{Vertical disk structure}\label{secver}

Since we wish to investigate the radial and vertical transport of tracers in 
protoplanetary disks as well as its feedback on the disk structure, it is 
necessary to resolve the vertical structure of the disk. For this purpose we
choose the $1$+$1$-dimensional approximation for the calculation of
the disk structure (e.g. Lin \& Papaloizou~\cite{lin85}). In this approximation
the vertical stratification is calculated separately for each grid point of the
radial one-zone model (see Sect.~\ref{secrad}) by assuming a hydrostatic
structure in the vertical direction. Therefore we accept some deviation of the
results of the $1$+$1$-dimensional model calculation from that of the exact
$2$-dimensional hydrodynamic calculation, but avoid the numerical complexity of
the latter.

\subsection{Vertical disk equations}

In the following the set of equations for the vertical disk structure in the 
$1$+$1$-dimensional approximation at a certain radius $r$ is given. 

We define a $z$-dependent column density: 
\begin{equation}\label{sigmaz}
\sigma(z) = \int_{z}^{\infty} \rho(z') \,dz' \ , 
\end{equation}
where $\rho(z)$ is the density at the height $z$ above the midplane at the given
radius $r$.  Note that $\sigma(z)=\frac12\Sigma(r)$ for $z \rightarrow 0$ with
this definition. The value of $\Sigma(r)$ is taken from the one-zone model.
The differential form of Eq.~(\ref{sigmaz}) is 
\begin{equation}\label{dsigmadz}
\frac{\partial \sigma}{\partial z} = - \rho \ . 
\end{equation}

The pressure stratification is determined by 
\begin{equation}\label{dpzdz}
\frac{1}{\rho}\frac{\partial P}{\partial z} = -\Omega^2{z} - 
  4\pi G \left( \frac{1}{2}\Sigma - \sigma(z) \right) \ , 
\end{equation}
where $P$ denotes the pressure. The second term on the r.h.s. of
Eq.~(\ref{dpzdz}) is the vertical acceleration of a self-gravitating infinite
slab (Paczy\'nski \cite{pac78}). The (vertical) self-gravity of the disk is
taken into account since in Paper V it has been found that the self-gravity
significantly modifies the disk structure for disk masses above $\sim 0.01\,
M_{\ast}$. Since in the present model calculations we choose an initial disk
mass of $M_{\rm disk} = 0.2\,M_{\ast}$, the self-gravity of the disk has to
be considered. 

The generation of heat in the disk is assumed to be caused exclusively by 
viscous friction. Thus the vertical gradient of the energy flux $F_{z}$ is given
by (e.g. Lin \& Papaloizou~\cite{lin85}) 
\begin{equation}\label{dfzdz}
\frac{\partial F_{z}}{\partial z} = \frac{9}{4} \Omega^2 \nu \rho \ . 
\end{equation}
Note that ${\dot E}_{\nu} = F_{z}(z \rightarrow \infty)$ holds according to 
Eq.~(\ref{dote}). Further note that irradiation
of the disk by the star is neglected in Eq.~(\ref{dfzdz}). 

The solution of Eq.~(\ref{dfzdz}) requires the specification of the vertical 
dependency of the viscosity $\nu$. However, the mechanism of turbulence in
disks as the main source of viscosity is a matter of much debate so far (see
e.g. Richard \& Davis~\cite{ric04}; Johansen et al. \cite{joh05,joh06}; Turner
et al. \cite{tur06}; Pavlyuchenkov \& Dullemond \cite{pav07} and references
therein). Due to the lack of
a precise knowledge of the vertical variation of $\nu$, the viscosity at point
($r$,$z$) is  set equal to the value which follows from the one-zone
approximation at $r$ (Eq.~(\ref{nubeta})), i.e., the viscosity is set
vertically constant. 

The heat is assumed to be transported solely by radiation. Heat transport by
convection is neglected since it is inefficient at low temperatures. In this
case, the vertical gradient of the temperature $T(z)$ is given by (e.g. Lin 
\& Papaloizou~\cite{lin85}) 
\begin{equation}\label{dtzdz}
\frac{\partial\, T(z)}{\partial z} = -\frac{3}{16}\, 
  \frac{ \kappa_{\rm R} \,\rho\, F_{z} }{ \sigma T^3(z) } \ . 
\end{equation}
The Rosseland mean opacity at each vertical location is calculated as in the 
radial model (Eqs.~(\ref{kafor}) -- (\ref{karp})). 

The set of vertical disk equations is closed by the equation of state for an 
ideal gas 
\begin{equation}\label{rhoz}
\rho = \frac{P}{c_{\rm s}^2} \ . 
\end{equation}

\subsection{Solution of the set of vertical disk equations}\label{secsolver}

The set of equations for the vertical disk structure (\ref{dsigmadz}) -- 
(\ref{rhoz}) constitutes a two-point boundary value problem. 
We define the top of the atmosphere at the height $z_0$ where the vertical 
optical depth is close to zero. 
According to the Eddington-Barbier approximation (e.g. Mihalas~\cite{mih78}) 
the temperature at $\tau = 0$ is determined by 
\begin{equation}\label{t0}
T_{0}^{4} = \frac{1}{2}\,T_{\rm eff}^{4} + T_{\rm cloud}^{4} \ . 
\end{equation}
The column density at $z_0$ has a very small value. 
In the model calculations we assume for $\sigma_{0}$ a fixed value of 
\begin{equation}\label{sigma0}
\sigma_{0} = \frac{1}{2} \cdot 10^{-6}\, \Sigma \ . 
\end{equation}
The energy flux at $z_0$ follows from Eq. (\ref{dfzdz}) to be 
\begin{equation}\label{f0}
F_{0} = \frac{9}{4} \Omega^2 \nu \left( \frac{1}{2}\Sigma - \sigma_{0} \right) 
  \ . 
\end{equation}
Finally the pressure at $z_0$ has to be determined. By assuming an isothermal
stratification for small optical depths Eq.~(\ref{dpzdz}) can be integrated. 
Its solution is 
\begin{equation}\label{p0}
P_{0} = \sqrt{\frac{2}{\pi}} \frac{ c_{0}^2 \,\sigma_{0} }{ h_{0} } 
  \frac{1}{1 - \Phi(x_{0})} \,e^{ \displaystyle - x_{0}^2 }
\end{equation}
with 
\begin{displaymath}\label{x0h0c0}
x_{0} = \frac{ 2 \pi G \,\sigma_{0}\, h_{0} }{ c_{0}^2 } + 
  \frac{1}{\sqrt{2}} \frac{z_{0}}{h_{0}} \ , \quad 
h_{0} = \frac{c_{0}}{\Omega} \ , \quad 
c_{0} = \sqrt{ \frac{ k_{\rm B} T_{0} }{ \mu_{0} m_{\rm H} } } \ . 
\end{displaymath}
$\Phi$ denotes the error function and $\mu_{0}$ the mean molecular weight at 
$z_0$ which is set to $7/3$.\footnote{
The temperature at $z_0$ is sufficiently low that the matter is in molecular
form at any radial position at the top of the atmosphere, and hence $\mu_{0}$ is close 
to $7/3$ for the standard cosmic element mixture.} 

Starting with an initial guess of the height $z_0$, the set of vertical disk equations (\ref{dsigmadz}) -- (\ref{rhoz}) is solved by a Runge-Kutta method of $5$th order (Press et al.~\cite{pre92}) from $z_0$ down to the midplane of the
disk at $z = 0$. The Runge-Kutta solver is equipped with a step size control
which keeps the number of vertical grid points small. In general, the column
density  $\sigma(z)$ at  $z = 0$ found by this procedure is not equal to the
midplane value $\Sigma / 2$. Then $z_0$ is iterated by a shooting method until
$\sigma(z=0)=\Sigma / 2$ within an accuracy of the vertical model of 
$\Delta_{\rm vert}=5 \cdot 10^{-5}$.


\section{Velocity Field}\label{secvel}

Finally, the velocity field in the $2$-dimensional disk model has to be
specified. We will show in this paper that the flow field in the disk affects
the radial mixing of tracers in the disk, exceeding the contribution of
turbulent diffusion. Therefore the actual flow field in the protoplanetary disk
has to be determined.

\subsection{Velocity field of the one-zone approximation}\label{secvelone}

The velocity field of the one-zone approximation is given by Eq.~(\ref{vr}). 
The main feature of this velocity field is that there exists a characteristic 
radial position $r_{x}$ where the sign of the radial velocity $v_{r}$ changes:

\begin{itemize}
 
\item Inward of $r_{x}$, $v_{r}$ is negative, i.e., the flow is directed inward. 
\item Outward of $r_{x}$, $v_{r}$ is positive, i.e., the flow is directed
outward.

\end{itemize} 
In model calculations of the solar nebula $r_{x}$ is typically located at a few 
or a few tens of ${\rm AU}$ and slowly changes with time (e.g. Ruden \& 
Lin~\cite{rud86}; Paper II). This is far outside of the chemical active zone
of the disk which is located  inside of about $1\,{\rm AU}$. Hence in one-zone
models of the solar nebula, outward radial mixing of tracers from the chemical
active zone to the outer solar system is only possible by turbulent diffusion
(Paper II; Paper~V).

\subsection{Meridional velocity field}\label{secmervel}

The real velocity field in disks is more complex than those of the one-zone 
approximation. Urpin~(\cite{urp84}) considered higher order terms in the flow
equations of the disk and found a characteristic flow pattern: 
\begin{itemize}

\item close to the disk midplane, the flow is directed outward whereas
\item at elevated heights above the midplane the flow is directed inward.

\end{itemize}
This type of {\it meridional flow field} has been confirmed to exist by
$2$-dimensional hydrodynamic calculations by Kley \& Lin~(\cite{kle92}),
R\'o\.zyczka et al.~(\cite{roz94}), Tscharnuter \& Gail (\cite{Tsa07}). Regev
\& Gitelman~(\cite{reg02}) conclude that the meridional flow field is a dynamical
phenomenon that is of universal occurrence in accretion disks.\footnote{In the $3$-dimensional
extension the flow field of accretion disks should additionally show large
vortices induced by the baroclinic instability (Klahr \& 
Bodenheimer~\cite{kla03}; Klahr~\cite{kla04}). 
} 

In the case of disks with $\alpha$-viscosity, Takeuchi \& Lin~(\cite{tak02}) 
derived analytic expressions for the radial and azimuthal velocity $v_{r}$ and 
$v_{\phi}$, respectively. In their work $v_{r}$ was calculated in first order
of the small perturbation parameter $\varepsilon \sim \frac{h_{\rm s}}{r} \sim
\frac{c_{\rm s}}{v_{\rm K}}$. $v_{\phi}$ is calculated in second order of 
$\varepsilon$ (which is the lowest non-vanishing order). Keller~(\cite{kel03}) and Paper VI
obtained the same result as Takeuchi \& Lin~(\cite{tak02}) and additionally
obtained the vertical velocity $v_{z}$ in $1$st order of $\varepsilon$. 

In the following the meridional velocity field for disks with $\beta$-viscosity
is derived in the lowest non-vanishing order of $\varepsilon$. According to
Takeuchi \& Lin~(\cite{tak02}) and Keller~(\cite{kel03}) the velocity field in
the stationary limit is given by 
\begin{equation}\label{vphi2d}
v_{\phi} = \left[ 1 - \frac{3}{4} \left( \frac{z}{r} \right)^2 + \frac{r^2}
  { 2 G M_{\ast} \rho } \frac{\partial P}{\partial r} \right] v_{\rm K} \ , 
\end{equation}
\begin{equation}\label{vr2d}
v_{r} = \left[ \frac{1}{r\rho} \frac{\partial}{\partial r} r^3 \rho\, \nu 
  \frac{\partial}{\partial r} \frac{v_{\phi}}{r} + \frac{r}{\rho} 
  \frac{\partial}{\partial z} \rho\, \nu \frac{\partial}{\partial z} v_{\phi} 
  \right] \frac{2}{v_{\rm K}} \ , 
\end{equation}
\begin{equation}\label{vz2d}
v_{z} = - \frac{1}{r\rho} \int_{0}^{z} 
  \frac{\partial}{\partial r} r \rho\, v_{r} \ . 
\end{equation}
As mentioned above, the lowest order, no-vanishing deviation of $v_{\phi}$ from
Keplerian rotation is of second order in $\varepsilon$. $v_{r}$ as well as 
$v_{z}$ are of first order in $\varepsilon$. To simplify Eqs.~(\ref{vphi2d}) --
(\ref{vz2d}), the density distribution of the stationary $1$+$1$-dimensional disk
model is used, i.e. (e.g. Pringle~\cite{pri81})\footnote{
In Eq.~(\ref{rho2d}) the self-gravity of the disk is neglected. It would be an
interesting task to include the disk's self-gravity in the calculation of the
meridional velocity field. } 
\begin{equation}\label{rho2d}
\rho(r,z) = \rho_{\rm c}(r) \,e^{ - z^2 / ( 2 H^2 ) } \ , 
\end{equation}
where 
\begin{equation}\label{rhocr}
\rho_{\rm c}(r) = \frac{|{\dot M}|}{ 6 \pi \nu H } \left( 1 - 
  \sqrt{ \frac{R_{\ast}}{r} } \right)
\end{equation}
is the midplane density and 
\begin{equation}\label{hs2d}
H = \sqrt{\frac{\pi}{2}} \frac{c_{\rm s}}{\Omega_{\rm K}}
\end{equation}
the scale height. 
$R_{\ast}$ denotes the radius of the protostar. 
Finally a radial power law for the midplane temperature is assumed, 
\begin{equation}\label{tcr2d}
T_{\rm c}(r) \propto r^{-\lambda} \ , 
\end{equation}
where $\lambda$ is the radial temperature exponent\footnote{
In principle, this assumption is not needed. The quantity $\lambda$ in all
equations for the velocity can simply be interpreted as the local value of
$\lambda=-$d\,$\ln T\,$/\,d\,$\ln r$.
The impact of the vertical temperature structure on the meridional flow field 
would be also of interest. However, this is not considered in the present 
work.
}.
For these density and temperature profiles, the velocity field in $\beta$-disks
is as follows: 
\begin{equation}\label{vphi2db}
v_{\phi}(r,z) = \left[ 1 - \frac{1}{4} \left( 4 + \lambda - f(r) \right) 
  \frac{c_{\rm s}^2}{v_{\rm K}^2} - \frac{1}{4} \lambda \frac{z^2}{r^2} \right] 
  v_{\rm K}
\end{equation}
\begin{equation}\label{vr2db}
v_{r}(r,z) = \beta \left[ 3 - \frac{5}{2} \lambda - \frac{3}{2} f(r) - 
  \frac{1}{2} ( 9 - 5 \lambda ) \left( \frac{z}{H} \right)^2 \right] v_{\rm K}
\end{equation}
\begin{equation}\label{vz2db}
v_{z}(r,z) = \beta \left[ C_1 \left( \frac{z}{r} \right) - C_2 \left( 
  \frac{z}{H} \right)^3 \right] v_{\rm K}
\end{equation}
with the factor 
\begin{displaymath}\label{fr}
f(r) = \left( \sqrt{ \frac{r}{R_{\ast}} } - 1 \right)^{-1}
\end{displaymath}
and coefficients 
\begin{displaymath}\label{c1}
C_1 = \frac{1}{4} \left[ ( 3 - \lambda ) ( 6 - 5 \lambda ) - ( 18 - 8 \lambda ) 
  f(r) \right]
\end{displaymath}
\begin{displaymath}\label{c2}
C_2 = \frac{1}{4} ( 3 - \lambda ) ( 9 - 5 \lambda ) \ . 
\end{displaymath}
To get a first impression of the meridional flow field see Fig.~\ref{fig2Dv4}b.

As in case of disks with $\alpha$-viscosity (Takeuchi \& Lin~\cite{tak02};
Paper VI) for the meridional velocity field of $\beta$-disks 
(Eqs.~(\ref{vphi2db}) -- (\ref{vz2db})) there exists a characteristic height 
$z_{\lambda}$ where the radial velocity changes its sign: 
\begin{equation}\label{zlambda}
\frac{z_{\lambda}}{H} = \sqrt{ \frac{ 6 - 5 \lambda -3 f(r) }
  { 9 - 5 \lambda } } \ ,
\smallskip
\end{equation}
provided the quantity under the square root is positive, which either requires
$\lambda< \frac{6}{5}$ (if $f(r)\ll1$) or $\lambda > \frac{9}{5}$. The latter
is not realised in accretion disks models calculated with reasonable dust
opacities.

If $\lambda < \frac{6}{5}$, which is satisfied in most parts of the disk (see
Sect.~\ref{secflofie} and Fig.~\ref{fig2Dv4}(b)), $v_{r}$ is negative above
$z_{\lambda}$, i.e., the flow is directed inward, whereas $v_{r}$ is positive
below $z_{\lambda}$, i.e., the flow is directed outward. Otherwise, in the case
$\frac{6}{5}<\lambda < \frac{9}{5}$, $v_{r}$ is negative at any height, and the
disk matter (the mixture of gas and small particles) moves inward at all heights.

The factor $f(r)$ affects the flow pattern only in the very inner parts of the 
disk ($f(r \rightarrow \infty) = 0$). As a consequence the location of
$z_{\lambda}$, and therefore the flow pattern of the disk, essentially depends
on the radial temperature exponent $\lambda$. 

In the model calculations of Paper VI, $\lambda$ has been chosen as a fixed 
parameter and set constant throughout the disk. In the present work $\lambda$
is calculated locally, i.e., at each grid point from the radial temperature
profile by a $3$-point interpolation. The interpolated value of $\lambda$ is
used to calculate the $2$-dimensional velocity field of the disk
($v_{r}$,$v_{z}$) given by Eqs.~(\ref{vr2db}), (\ref{vz2db}).

A vector transformation is implemented in the code for transforming the 
$2$-dimensional velocity field ($v_{r}$,$v_{z}$) into polar coordinates 
($v_{r}'$,$v_{\theta}$) since polar coordinates  are used to calculate the
transport of tracers (cf. Eq.~(\ref{dcijdtpol})). 

Note int Eq.~(\ref{vphi2db}) that the angular velocity $v_{\phi}$ deviates from
the Keplerian velocity $v_{\rm K}$. In particular it varies with height $z$.
However, the value of $v_{\phi}$ from Eq.~(\ref{vphi2db}) is not required in the
present $2$-dimensional model.

\section{Numerical treatment}\label{secnum}

\subsection{Initial and boundary conditions}

The boundary conditions are chosen as follows: 
\begin{itemize}
\item At the outer boundary ($r_{\rm o} = 200\,{\rm AU}$) we set $\Sigma = 0$ 
  (no-torque condition). 
\item At the inner boundary ($r_{\rm i} = 0.096\,{\rm AU}$) we choose the 
  quasi-stationary boundary condition which has been introduced in Paper V. 
  The quasi-stationary boundary condition removes the problem of the unphysical 
  behaviour of $\Sigma$ in the inner zone of the disk which appears in the case 
  of the no-torque condition $\Sigma = 0$ and ensures a smooth and physically 
  more realistic radial distribution of the surface density (as well as other 
  quantities) in the inner disk zone. 
\item The concentrations $c_{i,j}$ at the outer boundary are chosen such that 
  the silicates are amorphous ($f_{\rm cry,for} = f_{\rm cry,ens} = 0$) and all 
  carbon that is not bound in ${\rm CO}$ is condensed into solid carbon 
  ($f_{\rm car} = 0.6$). 
\item The $c_{i,j}$ at the inner boundary are chosen such that the silicates 
  are completely crystalline ($f_{\rm cry,for} = f_{\rm cry,ens} = 1$) and 
  solid carbon is completely oxidised ($f_{\rm car} = 0$). 
\item At the upper boundary of the $2$-dimensional polar grid 
  ($\theta_{1} = 4.01\degr$) the dust is assumed to have interstellar 
  properties as is assumed for the outer boundary. 
  Hence, the $c_{i,j}$ at the upper boundary are chosen such that the silicates 
  are amorphous ($f_{\rm cry,for} = f_{\rm cry,ens} = 0$) and a fraction of
  $f_{\rm car} = 0.6$ of the total carbon is assumed to be bound in solid carbon
  grains, the remaining fraction being in ${\rm CO}$. 
\item For the upper boundary as well as the inner and outer boundary 
  homogeneous von Neumann boundary conditions are chosen for the tracres, i.e.,
  we set $\frac{\partial c_{i,j}}{\partial \theta} = 0$ at the upper boundary
  and $\frac{\partial c_{i,j}}{\partial r'} = 0$ at both lateral boundaries. 
  This choice prevents any flux of matter across the boundaries. 
\item At the midplane of the disk symmetry conditions are applied. 
\end{itemize}

The initial conditions for the one-zone model are chosen as in Paper V for the 
model with solar abundance, i.e., the initial radial distributions of $\Sigma$ 
and $c_{i,{\rm car}}$ are calculated from a stationary model 
($\partial\Sigma/\partial t = 0$ and $\partial c_{i,{\rm car}}/\partial t = 0$) 
in which the silicates are assumed to be unaltered amorphous ISM grains. 

The $c_{i,j}$ at the position ($r'$,$\theta$) of the $2$-dimensional polar grid 
initially are set equal to the midplane value $c_{i,j}(r=r')$, i.e., the 
$c_{i,j}$ initially are set constant in $\theta$-direction.

\begin{figure}[t]
\setlength{\unitlength}{1 true cm}
\begin{center}
\begin{picture}(15,17.3)
%
\put(0,16.5){\framebox(7.0,0.6){Parameters ($\beta$, $M_{\rm disk}$, 
  $J_{\rm disk}$, $\dots$)}}
\put(0,15.5){\framebox(7.0,0.6){Initial model ($\Sigma(r,t=0)$, 
  $c_{i,j}(r',\theta,t=0)$)}}
\put(0,14.5){\framebox(7.0,0.6){Set of radial disk Eqs.~(\ref{omega}) -- 
  (\ref{vr})}}
\put(0,13.5){\framebox(7.0,0.6){NR iteration of ($T_{\rm c}$,$\mu$) up to 
  $\Delta = 9{\cdot}10^{-6}$}}
\put(0,12.5){\framebox(7.0,0.6){Solution of Eq.~(\ref{dsigmadt}) for 
  $\Sigma(r)$}}
\put(0,11.5){\framebox(7.0,0.6){Solution of the set of Eqs.~(\ref{dcijdt}) for 
  the $c_{i,j}(r)$}}
\put(0,10.5){\framebox(7.0,0.6){New $f_{\rm cry,for}(r)$, $f_{\rm cry,ens}(r)$ 
  and $f_{\rm car}(r)$}}
\put(0,9.5){\framebox(7.0,0.6){Fixpoint iteration of $\Sigma(r)$ up to 
  $\Delta = 10^{-5}$}}
\put(0,8.5){\framebox(7.0,0.6){Set of vertical disk Eqs.~(\ref{dsigmadz}) -- 
  (\ref{rhoz})}}
\put(0,7.5){\framebox(7.0,0.6){Iterating up to an accuracy of 
  $\Delta = 5 \cdot 10^{-5}$ for $z_0$}}
\put(0,6.5){\framebox(7.0,0.6){Interpolation from the ($r$,$z$) grid onto the 
  $(r',\theta)$ grid}}
\put(0,5.5){\framebox(7.0,0.6){Solution of the set of Eqs.~(\ref{dcijdtpol}) 
  for the $c_{i,j}(r',\theta)$}}
\put(0,4.5){\framebox(7.0,0.6){Interpolation from the ($r$',$\theta$) grid onto 
  the $(r,z)$ grid}}
\put(0,3.5){\framebox(7.0,0.6){New $f_{\rm cry,for}(r,z)$, 
  $f_{\rm cry,ens}(r,z)$ and $f_{\rm car}(r,z)$}}
\put(0,2.5){\framebox(7.0,0.6){Fixpoint iteration of $T(r',\theta)$ up to 
  $\Delta = 5 \cdot 10^{-3}$}}
\put(0,1.5){\framebox(7.0,0.6){New stellar mass $M_{\ast}$ from 
  Eqs.~(\ref{dotm}) and (\ref{mstar})}}
\put(0,0.5){\framebox(7.0,0.6){Next time step}}
\thicklines
\multiput(3.5,1.5)(0,1){16}{\vector(0,-1){0.4}}
\put(7,13.8){\line(1,0){0.5}}
\put(7.5,13.8){\line(0,1){0.8}}
\put(7.5,14.6){\vector(-1,0){0.5}}
\put(7,9.8){\line(1,0){1.0}}
\put(8,9.8){\line(0,1){5.0}}
\put(8,14.8){\vector(-1,0){1.0}}
\put(7,7.8){\line(1,0){0.5}}
\put(7.5,7.8){\line(0,1){0.9}}
\put(7.5,8.7){\vector(-1,0){0.5}}
\put(7,2.8){\line(1,0){1.0}}
\put(8,2.8){\line(0,1){6.1}}
\put(8,8.9){\vector(-1,0){1.0}}
%
\put(7,0.8){\line(1,0){1.5}}
\put(8.5,0.8){\line(0,1){14.2}}
\put(8.5,15.0){\vector(-1,0){1.5}}
\end{picture}
\caption{
Flow chart of the code. The arrows to the right of the diagram denote the
iteration loops and the time loop, respectively. The inner two loops in the
upper half refer to the one-zone model and the inner two loops in the lower
half to the $2$-dimensional model. $\Delta$ denotes the accuracy of the
iteration. `NR' is an abbreviation for Newton-Raphson. The set of radial disk
Eqs.~(\ref{omega}) -- (\ref{vr}) is solved by a coupled Newton-Raphson method
for the midplane temperature $T_{\rm c}$ and the mean molecular weight $\mu$. 
For details see text.}\label{flow}
\end{center}
\end{figure}
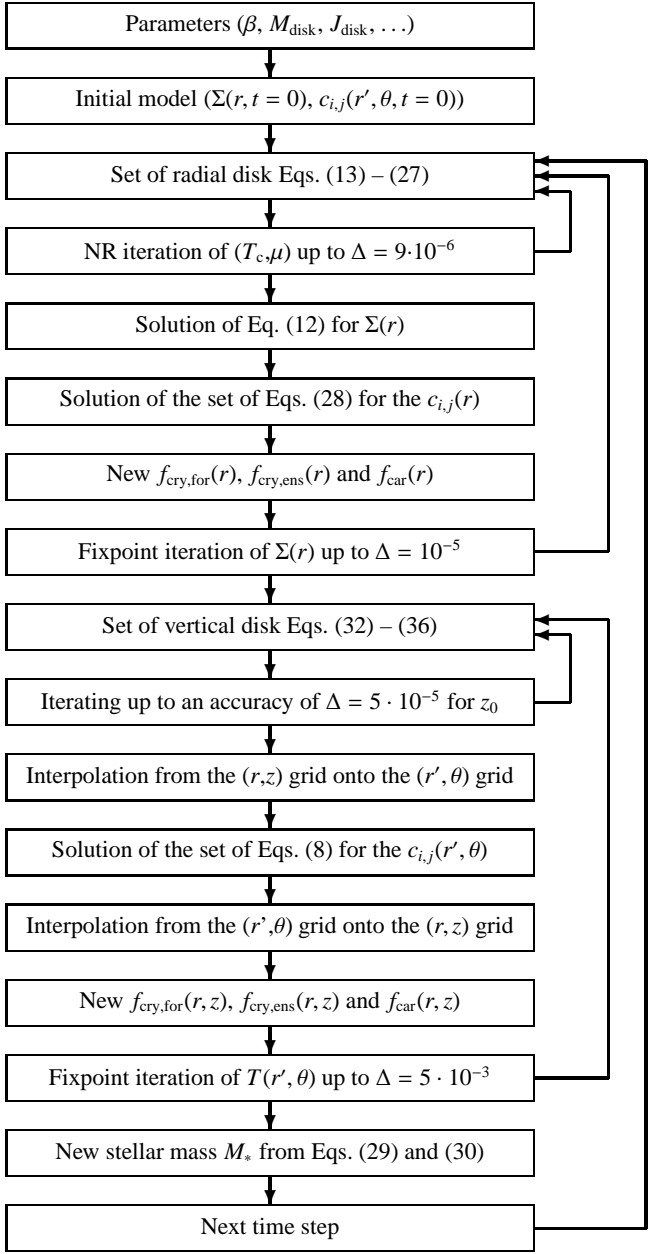

\subsection{2-dimensional grids}

In the present model the disk structure is calculated in cylindrical coordinates
($r$,$z$) ($1$+$1$-dimensional approximation) whereas the tracer transport is 
calculated in polar coordinates ($r'$,$\theta$). For a self-consistent
computation of the tracer transport and the disk structure an interpolation of
the relevant physical quantities between both grids is applied. This is done by
a $2$D bilinear interpolation (Press et al.~\cite{pre92}). 

The radial grid extends from $0.096\,{\rm AU}$ to $200\,{\rm AU}$ and consists 
of $332$ logarithmically equidistant grid points ($100$ grid points per decade).
The vertical grid size depends on the radial position and is variable in time 
since we adopt a step size control for $z$ in the calculation of the vertical
disk structure. In the chemically active zone of the disk the number of vertical
grid points usually is larger than $100$ (maximum $\sim 170$ grid points)
whereas it drops below $100$ in the cool outer disk regions. The total number of
grid points of the cylindrically symmetric grid adds up to about $35\,000$. 

For the polar grid an apex angle of $\theta_1 = 4.01\degr$ has been chosen. 
This choice ensures that the polar grid is embedded in the cylindrically symmetric
grid at any time of the model calculations. The polar angle $\theta$ is
discretised in $L = 51$ equidistant grid points $\theta_l$ between the upper
boundary $\theta_1 = 4.01\degr$ and the midplane $\theta_L = 0\degr$. Note that
the polar grid defines the 'domain of transport`, i.e., the domain where the
transport of tracers by advection and diffusion occurs during the model
calculations. Beyond the polar grid no transport takes place ($v_{r} = 
v_{z} = 0$, $D = 0$). We did test calculations with $L = 101$ polar grid points
as well as with an apex angle of $\theta_1 = 2.01\degr$ and found no significant
deviations from the standard model with $L = 51$ and $\theta_1 = 4.01\degr$.

\subsection{Numerical Solution}

The flow chart of the code of the present disk model is shown in 
Fig.~\ref{flow}. The model calculations are performed as follows: 

First the radial disk structure in the one-zone approximation is calculated 
with standard methods (inner two loops in the upper half of Fig.~\ref{flow}; 
see Sect.~\ref{secrad} and cf. Papers II and V). The equation for $\Sigma$,
Eq. (\ref{dsigmadt}), is solved fully implicit.

Secondly, the vertical disk structure is computed (innermost loop in the lower 
half of Fig.~\ref{flow}; see Sect.~\ref{secver}). In this way we obtain the
disk structure in cylindrical coordinates ($r$,$z$). 

Thirdly, the $2$-dimensional transport of tracers is calculated. This is done by
\begin{enumerate}
\item Interpolating relevant quantities ($T$, $D$, $n$, $n_{\rm OH}$, $v_r$,
  $v_z$)  from the cylindrically symmetric grid ($r$,$z$) onto the polar grid
  $(r',\theta)$, 
\item solving the set of $2$D tracer equations~(\ref{dcijdtpol}) in polar 
  coordinates (see Sect.~\ref{sectra}), 
\item calculating the degrees of crystallisation of the silicates, 
  $f_{\rm cry,for}$ and $f_{\rm cry,ens}$, as well as the degree of 
  condensation of ${\rm C}$ in solid carbon, $f_{\rm car}$, from the 
  $c_{i,j}(r',\theta)$ (see Eqs.~(\ref{fcryfor}) -- (\ref{fcar})), and 
\item interpolating $f_{\rm cry,for}$, $f_{\rm cry,ens}$ and $f_{\rm car}$ from
  the polar grid $(r',\theta)$ onto the cylindrically symmetric grid ($r$,$z$). 
\end{enumerate}
To calculate the tracer transport self-consistently with the disk structure the 
temperature is iterated globally by a fixpoint iteration up to an accuracy of 
$\Delta_{\rm gl} = 5 \cdot 10^{-3}$ (outer loop in the lower half of 
Fig.~\ref{flow}). The relative low accuracy of the global iteration is attributed to the errors which are introduced by the interpolation
procedure. We did test calculations without performing the global iteration and
found only small deviations from the standard model with $\Delta_{\rm gl}=
5\cdot 10^{-3}$ which are not relevant for the final results.

Finally, the stellar mass is updated (see Eqs.~(\ref{dotm}) and (\ref{mstar})) 
before the next time step is executed (outermost loop in Fig.~\ref{flow}).

\subsection{Model parameters}

\begin{table}
\begin{center}
\caption{
Parameters used for the calculation of the disk models.}\label{tabpara}

\begin{tabular}{l@{\extracolsep\fill}cr@{\hspace{.1cm}}l}
\hline\hline
\noalign{\medskip}
Initial stellar mass          & $M_{\ast,0}$     & $1$                 & 
  $M_{\sun}$ \\
Stellar effective temperature & $T_{\ast}$       & $4\,250$            & 
  ${\rm K}$ \\
Stellar luminosity            & $L_{\ast}$       & $5$                 & 
  $L_{\sun}$ \\
Stellar radius                & $R_{\ast}$       & $4.13$              & 
  $R_{\sun}$ \\
Inner disk radius             & $r_{\rm i}$      & $5\,R_{\ast}$       & \\
                              &                  &=0.096               &
   AU     \\
Outer disk radius             & $r_{\rm o}$      & $200$               & 
  ${\rm AU}$ \\
Molecular cloud temperature   & $T_{\rm cloud}$  & $20$                & 
  ${\rm K}$ \\
Initial disk mass             & $M_{\rm disk,0}$ & $0.2$               & 
  $M_{\sun}$ \\
Disk angular momentum         & $J_{\rm disk}$   & $10^{53}$           & 
  ${\rm g\,cm}^2\,{\rm s}^{-1}$ \\
Apex angle of the polar grid  & $\theta_1$       & $4.01$              & 
  $\degr$ \\
Viscosity parameter           & $\beta$          & $10^{-5}$           & \\
Number of polar   &  &  $332 \times 51$ \\
\quad grid points &  &= $16\,932$ & \\
Number of cylindrical&                  & $\sim 35\,000$      & \\
\quad  grid points & & \\
\noalign{\smallskip}
\hline
\end{tabular}
\end{center}
\end{table}
The model parameters for the disk models calculated in the present work are 
shown in Table~\ref{tabpara}. 
The stellar and disk parameters are chosen as in Paper V and are typical for 
solar like T Tauri stars and their surrounding accretion disks (see references 
given in Paper II). 
With respect to the choice of the value of the viscosity parameter $\beta$ the 
reader also is referred to Paper V and the discussion therein. 

The model calculations are initiated with a small time step of $10^{-8}\,{\rm
yr}$ which is increased slowly by an implemented time step control. 
The time step is limited to $50\,{\rm yr}$ to ensure numerical stability. 
We quit the model calculations at $10^{6}\,{\rm yr}$. The model with the above
standard parameters requires a CPU time of about two month on a P4 XEON 
$2.8\,{\rm GHz}$ computer.


\section{Results}\label{secres}

\subsection{Models 1DC and 1DP}\label{secmod1D}

Before we present the results of the $2$D model calculations we first compare 
two model calculations in the one-zone approximation which correspond to two 
different underlying disk geometries. 

In the first model, henceforth called model 1DC, the tracer transport is 
calculated in cylindrical coordinates ($r$,$z$) by assuming vanishing 
concentration gradients in the vertical direction, i.e. ,
$\frac{\partial c_{i,j}}{\partial z} = 0$. 
The transport-reaction equation in the one-zone approximation then is given by 
Eq.~(\ref{dcijdt}), i.e.,
\begin{equation}\label{dcijdtcyl}
\frac{\partial c_{i,j}}{\partial t} + v_{r} \frac{\partial c_{i,j}}{\partial r} 
  = \frac{1}{rn} \frac{\partial}{\partial r}\, rnD\, 
  \frac{\partial c_{i,j}}{\partial r} + \frac{R_{i,j}}{n} \ . 
\end{equation}
Model 1DC essentially equals the standard model with solar element mixture in 
Paper V. 

In contrast to model 1DC, the tracer transport in the second model, henceforth 
called model 1DP, is calculated in polar coordinates ($r'$,$\theta$) within a 
one-zone model. For this purpose we assume the concentration gradients in polar
direction to be negligible, i.e., $\frac{\partial c_{i,j}}{\partial \theta} = 0$. 
This assumption is almost identical to the assumption $\frac{\partial c_{i,j}}
{\partial z} = 0$ made for Eq.~(\ref{dcijdtcyl}). It is justified by the fact
that the disk model is based on the idea of a flat and flaring disk ($z \ll r$)
which has a small apex angle. As well it is justified by the results of the 
$2$D model calculations of the present work which show only slight concentration
gradients in the vertical (respectively polar) direction since the vertical
(polar) diffusion erases almost any vertical concentration gradient. 

With these basic settings the transport-reaction equation in $2$D polar 
coordinates, Eq.~(\ref{dcijdtpol}), transforms into 
\begin{equation}\label{dcijdtpolr}
\frac{\partial c_{i,j}}{\partial t} + v_{r}' 
  \frac{\partial c_{i,j}}{\partial r'} = \frac{1}{r'^2 n}\, 
  \frac{\partial}{\partial r'}\, r'^2 n D\, 
  \frac{\partial c_{i,j}}{\partial r'} + \frac{R_{i,j}}{n} \ . 
\end{equation}
Equation~(\ref{dcijdtpolr}) determines the tracer transport in the one-zone 
model 1DP. 

Both Eqs.~(\ref{dcijdtcyl}) and (\ref{dcijdtpolr}) determine the tracer
concentrations in the midplane of the disk where $r = r'$ but the 
geometry of the disk in both models is different. In model 1DC the disk is a
flat slab whereas in model 1DP the disk resembles an outward flaring slab. 
For this reason in both models the diffusion terms in the transport-reaction 
equations, Eqs.~(\ref{dcijdtcyl}) and (\ref{dcijdtpolr}), differ from each 
other. 
\begin{figure}[t]
\resizebox{1.0\hsize}{!}{\includegraphics{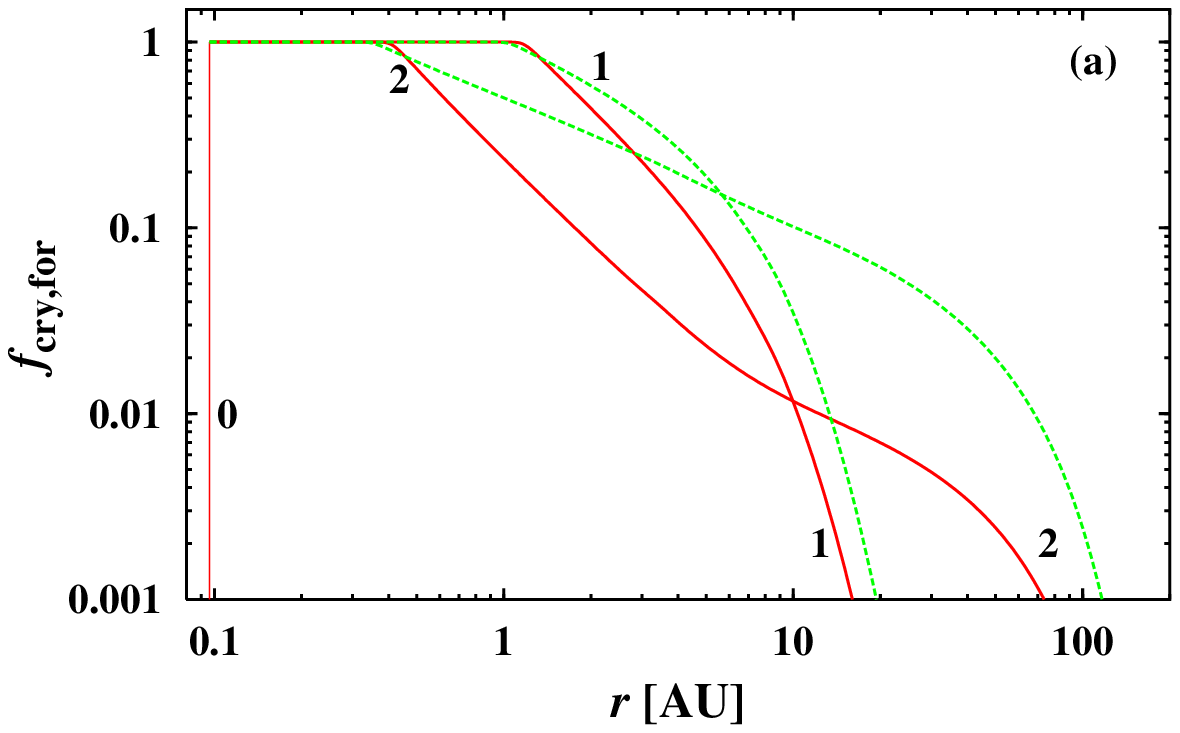}}

\resizebox{1.0\hsize}{!}{\includegraphics{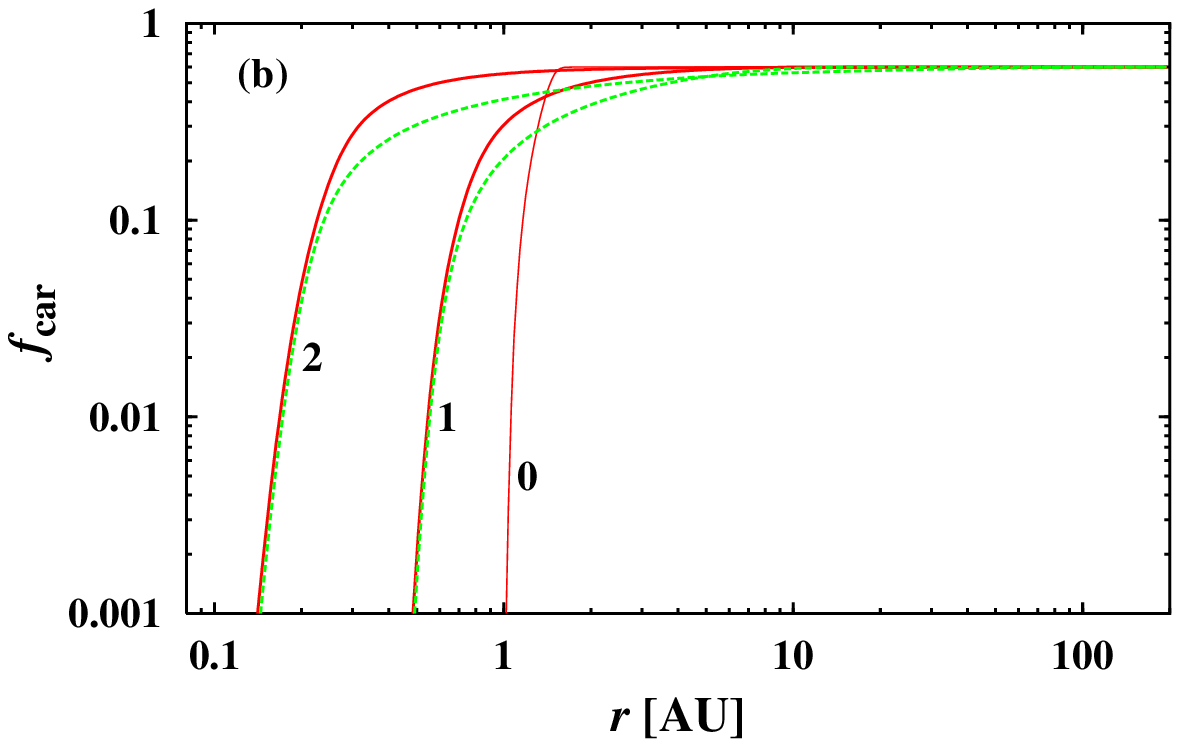}}
\caption{
Tracer transport in one-zone models with different geometries at times $t=0$ 
(0), $10^5$ (1) and $10^6\,{\rm yr}$ (2). The 'cylindrical` model 1DC is shown
with dashed lines and the 'polar` model 1DP with solid lines. 
{\bf (a)} Degree of crystallisation of forsterite $f_{\rm cry,for}$. 
{\bf (b)} Fraction of ${\rm C}$ condensed in solid carbon $f_{\rm car}$. 
}\label{figf1D}
\end{figure}

The radial velocity profile in both models is assumed to be that of the 
one-zone model which is given by Eq.~(\ref{vr}). 

Note that the radial density profiles $n(r)$ that enters in the diffusion terms of Eqs. (\ref{dcijdtpolr}) and (\ref{dcijdtpolr}) in both models is identically 
to the density profile of a flaring disk (cf. Eq.~(\ref{rhom})). However, to
compare the tracer transport in models 1DC and 1DP one actually has to apply a
density profile for model 1DC with {\it constant} scale height to account for
the cylindrical disk geometry in this model. In contrast, the polar disk geometry
in model 1DP represents to be a good approximation of the real flaring disk
geometry. We realised this inconsistency after the termination of the model
calculations. However, we refrain from recalculating model 1DC with a density
profile with constant scale height since we intend to compare the model
calculations of the present paper with those of the previous papers of this
series which computes the tracer transport as model 1DC, i.e., with cylindrical
disk geometry and 'flaring` $n(r)$. 

The dependence of the results for the tracer transport on the different assumed 
disk geometries is shown in Fig.~\ref{figf1D}. 

In Fig.~\ref{figf1D}a the degree of crystallisation of forsterite $f_{\rm cry,
for}$ at $t=0$ (0), $10^5$ (1) and $10^6\,{\rm yr}$ (2) is plotted for the
models 1DC (dashed lines) and 1DP (solid lines). It clearly can be seen that the
'polar` model 1DP leads to a substantially less efficient radial mixing than the
'cylindrical` model 1DC. In the region around $10\,{\rm AU}$ after $10^6\,
{\rm yr}$ of the disk evolution, $f_{\rm cry,for}$ in model 1DP is about one
order of magnitude lower than in model 1DC. This can be explained by the disk
geometry. In model 1DP the disk flares. As a result in model 1DP tracers are
diluted to a greater extent as they are mixed outward as compared to model
1DC where the geometry is similar to a tube with parallel walls. However, polar
coordinates obviously resemble more closely the real disk geometry than
cylindrical coordinates do. Thus, in former one-zone models of the solar nebula
that treat the tracer transport in cylindrical geometry (Paper II; Paper V;
Bockel\'ee-Morvan et al.~\cite{boc02}) radial mixing has been overestimated. As
a consequence the hypothesis of radial mixing that explains many interesting
properties of primordial solar system bodies hardly can be maintained for
previous one-zone models. For example the observed fraction of crystalline
silicates of more than $10\,\%$ of the bulk silicate in many comets (e.g. Hanner
et  al.~\cite{han94}; Crovisier et al.~\cite{cro97}; Hanner et al.~\cite{han97};
Yanamandra-Fisher \& Hanner~\cite{yan99}; Wooden et al. \cite{woo05,woo07})
hardly can be explained by one-zone models with a realistic polar disk geometry. 

A way out of this dilemma is provided by $2$-dimensional models with a
realistic description of the flow field in the disk (see Sect.~\ref{sec2dmoca}).

\begin{figure*}[t]
\resizebox{1.0\hsize}{!}
  {\put(2.0,8.5){}\includegraphics{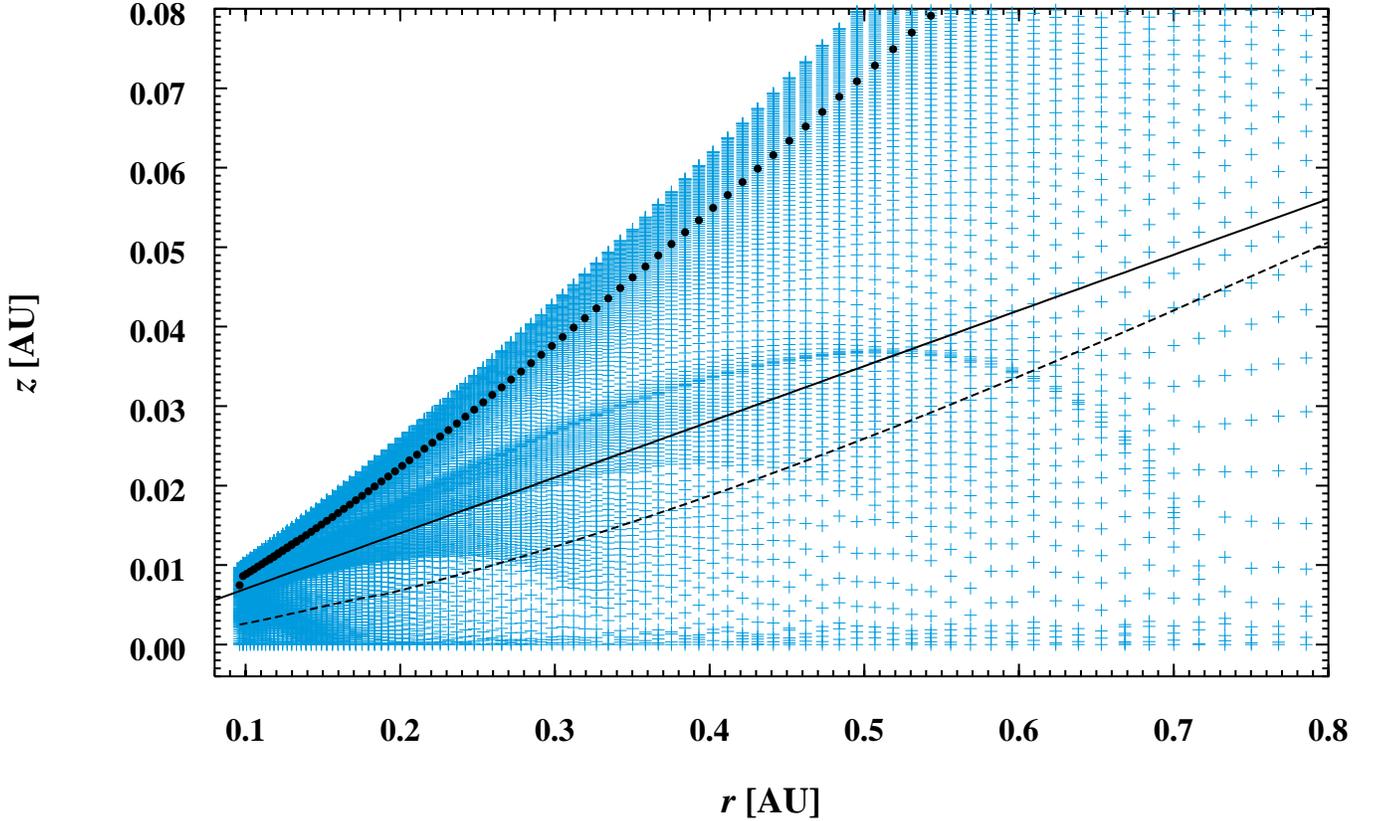}}
\caption{
$1$+$1$-dimensional grid of the standard model at $t = 0$ in the inner part of
the disk. Each cross marks a grid point. The dashed line shows the radial
profile of the pressure scale height $h_{\rm s}$ (Eq.~(\ref{hs})) and the
solid line the upper boundary of the $2$D polar grid. The big dots close to the
upper boundary of the grid mark the location of the photosphere ($\tau = 1$).
The arc-like structures mark the location of the vapourisation zones of strong
absorbers (for details see text). 
}\label{figzr0}
\end{figure*}

Carbon dust in model 1DP is mixed radially outward more efficiently than in
model 1DC (Fig.~\ref{figf1D}b). This can be explained by the dilution of the
products of carbon combustion in the outer parts of the disk (e.g. ${\rm CO}$,
${\rm CH}_4$; Gail~\cite{gai02}) which is more pronounced in the polar model
1DP than in the cylindrical model 1DC. This again is due to the geometric effect
described above for the radial mixing of the crystalline silicate grains. 

\subsection{1+1-dimensional grid}

In the next step we turn from the one-zone disk model to the 
$1$+$1$-dimensional disk model with $2$-dimensional transport of tracers. 

To give an impression of the $1$+$1$-dimensional cylindrically grid it is
plotted in Fig.~\ref{figzr0} at time $t = 0$ of the standard model 2DM for the 
innermost $0.8\,{\rm AU}$ of the disk and $z\ge0$. As been described in
Sect.~\ref{secsolver}, the solver for the disk vertical structure is equipped
with an automatic step size control keeping the number of vertical grid points
small. In particular the step size $\Delta z$ is small in such regions where
physical  quantities show large gradients. Such regions clearly can be seen in
Fig.~\ref{figzr0}. 

On the one hand the grid point density gets high close to the upper boundary 
of the vertical grid as a consequence of the large pressure and density 
gradients. 

On the other hand the grid point density is increased in the narrow zones of 
dust vapourisation/destruction since temperature and opacity are strongly
variable there. As a result, arc-like structures of enlarged grid point density
are formed which stand out from the ambient grid with low resolution. An example
is the arc-like structure that culminates at height $z \approx 0.035\,{\rm
AU}$ and extends out to $r \approx 0.7\,{\rm AU}$. This results from the
transformation of enstatite to forsterite. Above and to the right of the arc
enstatite and forsterite are both stable whereas beneath the arc enstatite is
unstable and forsterite is the only stable silicate. The enstatite-to-forsterite
transformation occurs at about $1\,280\,{\rm K}$. The two arcs culminating at
$z \approx 0.02$ and $0.01\,{\rm AU}$ display the vapourisation fronts of
forsterite ($T \sim 1\,360\,{\rm K}$) and solid iron ($T \sim 1\,380\,
{\rm K}$), respectively. The vapourisation of corundum which is the most
refractive dust species in the present model calculations occurs at $r \approx
0.12\,{\rm AU}$ and $T \sim 1\,755\,{\rm K}$ (midplane values at $t =
0$).

For comparison, in Fig.~\ref{figzr0} also the radial run of the pressure 
scale height $h_{\rm s}$ (Eq.~(\ref{hs}); dashed line), the upper boundary of 
the $2$D polar grid (solid line) and the location where the atmosphere gets 
vertically optically thin ($\tau = 1$; big dots) is shown. The $2$D polar grid
for tracer transport (not shown) is well embedded in the $1$+$1$-dimensional
cylindrically symmetric grid. Moreover, the $2$D polar grid is well located
within the vertically optically thick region of the disk. 

Note again that the $2$D polar grid defines the area of turbulent and advective
transport. Beyond these area no diffusive or advective transport is considered
in the present model calculations. 

\subsection{2-dimensional model calculations}\label{sec2dmoca}

In the final step we present the results of the 2-dimensional model 
calculations. For the purpose of investigating the influence of the disk`s
flow field on the tracer transport two model calculations with different flow
patterns are performed: 
\begin{itemize}
\item In the first model, henceforth called model 1DE, the flow field of the 
  one-zone model (Eq.~(\ref{vr})) is extended to the disk regions off the 
  midplane, i.e., at each polar grid point the polar radial velocity $v_{r}'$
  is chosen equal to the one-zone radial velocity $v_r$, $v_{r}'(r',\theta) =
  v_{r}(r=r')$, and the polar velocity $v_{\theta}$ everywhere is set to zero,
  $v_{\theta}(r',\theta) = 0$. Although this velocity field is $2$-dimensional
  we call it in the following the `extended one-zone' velocity field (1DE). 
\item In the second model, henceforth called model 2DM, the meridional velocity
  field is used which is calculated as described in Sect.~\ref{secmervel}. 
\end{itemize}
%

\subsubsection{Flow fields in models 1DE and 2DM}\label{secflofie}

\begin{figure*}[t]
\centerline{
\resizebox{0.9\hsize}{!}{\includegraphics{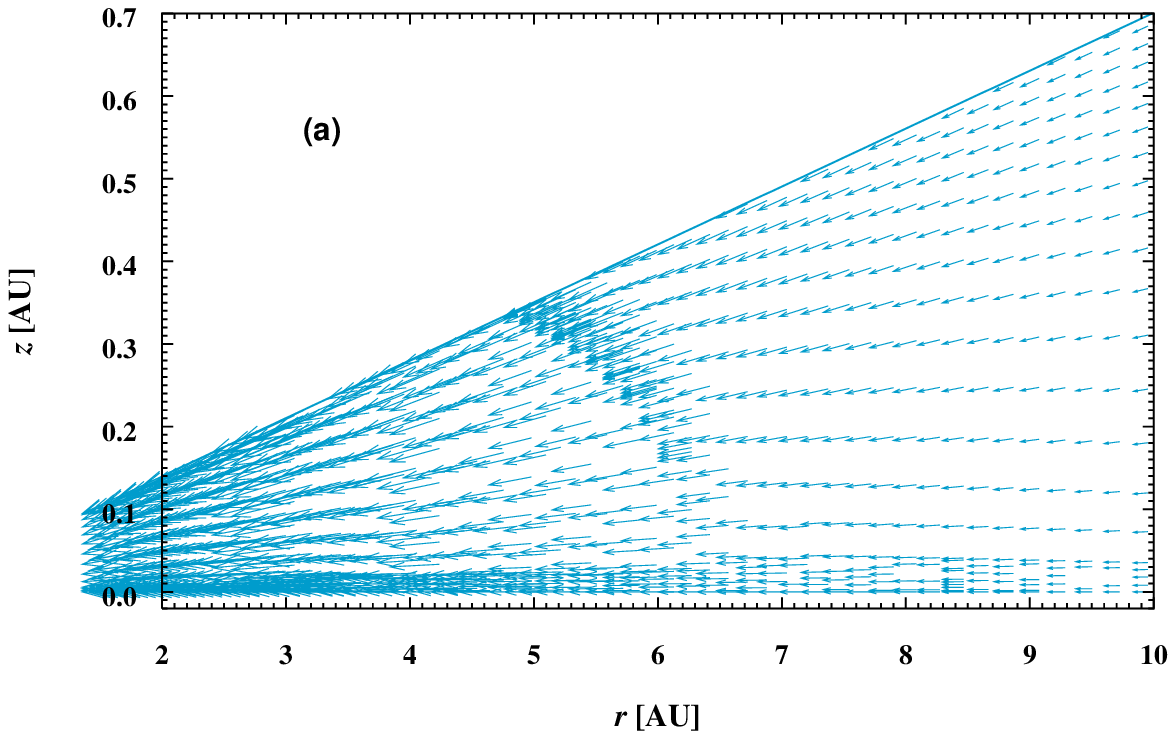}}
}
\centerline{
\resizebox{0.9\hsize}{!}{\includegraphics{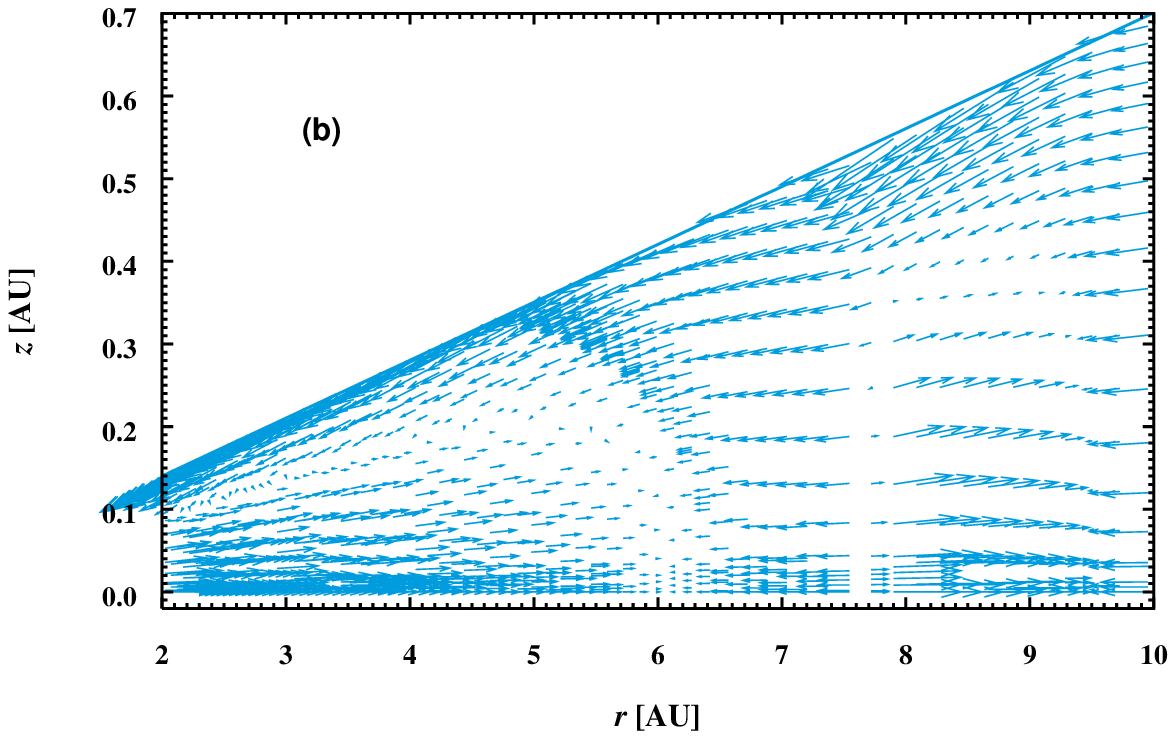}}
}

\caption{
$2$-dimensional flow field in the disk at $t = 10^5\,{\rm yr}$ between $2$ and 
$10\,{\rm AU}$. The solid line marks the upper boundary of the polar grid 
($\theta_1 = 4.01\degr$). {\bf (a)} Model 1DE. {\bf (b)} Model 2DM. 
The arrows show the direction of the flow; their length is proportional to the
velocity. For details see text. 
}\label{fig2Dv4}
\end{figure*}

To compare the extended one-zone velocity field and the meridional velocity
field the flow fields of both models 1DE and 2DM are plotted in
Fig.~\ref{fig2Dv4} at time $t = 10^5\,{\rm yr}$ in the region between $2$ and
$10\,{\rm AU}$. 

The one-zone velocity field (Fig.~\ref{fig2Dv4}a) is directed inward for most 
parts of the disk as it was mentioned in Sect.~\ref{secvelone}. The velocity
increases inward as the viscous torque increases with decreasing $r$. The
absolute value of the velocity at $4\,{\rm AU}$, e.g., is $20\,{\rm cm\,s}^{-1}$
whereas at the inner boundary it is $260\,{\rm cm\,s}^{-1}$. The feature with
large grid point density at about $6\,{\rm AU}$ marks the ice condensation
front. 

The meridional flow field (Fig.~\ref{fig2Dv4}b) shows a distinctly more complex
structure than the one-zone velocity field. The meridional flow pattern shows
large eddies which extend from both sides of the midplane of the disk to the
disk's surfaces. Particularly the structure of the flow field in the region of
the ice condensation front is of interest as the radial temperature
distribution shows a complex behaviour in that region. In detail the midplane
radial velocity $v_{r}'(r',\theta=0)$ shows the following radial dependence in
model 2DM at $t = 10^5\,{\rm yr}$: 
\renewcommand{\labelenumi}{(\roman{enumi})}
\begin{enumerate}
\item For $r \lesssim 6\,{\rm AU}$, $v_{r}$ is positive since for the radial 
  temperature exponent in this region $\lambda < \frac{6}{5}$ holds (cf. 
  Sect.~\ref{secmervel}). 
\item In the range of $6\,{\rm AU} \lesssim r \lesssim 7.5\,{\rm AU}$, $v_{r}$ 
  is negative.   There the opacity is a decreasing function of $r$ that
  results in a radial  temperature profile with $\lambda > \frac{6}{5}$. 
\item In the range $7.5\,{\rm AU} \lesssim r \lesssim 9.5\,{\rm AU}$, $v_{r}$ 
  again is positive as the condensation front of ice leads to a temperature 
  plateau, i.e., $\lambda < \frac{6}{5}$. 
\item In the range $9.5\,{\rm AU} \lesssim 22\,{\rm AU}$, $v_{r}$ for the same 
  reasons as in (ii) is negative. 
\item Finally, for $r \gtrsim 22\,{\rm AU}$, $v_{r}$ is positive since the 
  temperature adjusts to the ambient cloud temperature 
  ($T_{\rm cloud} = 20\,{\rm K}$), i.e., $\lambda < \frac{6}{5}$. 
\end{enumerate}
\renewcommand{\labelenumi}{(\arabic{enumi})}

Due to the accretion process the regions (i) -- (v) slowly move radially
inward. Additionally, the diffusive spreading of the disk causes a flattening
of the radial temperature profile which narrows the ranges (ii) and (iv) with
negative $v_{r}$ with time. Despite of this, the flow patterns (i) -- (v)
around the ice front persists for more than $10^6\,{\rm yr}$ of the disk
evolution. The chemical active zone where in particular annealing of
silicates and carbon combustion occurs, is always located within range (i). 

The polar velocity $v_{\theta}$ is more than one order of magnitude lower than 
$v_{r}'$ close to the midplane, and therefore plays no important role with 
regard to mixing processes in the disk. 

\begin{figure}[t]
\resizebox{1.0\hsize}{!}{\includegraphics{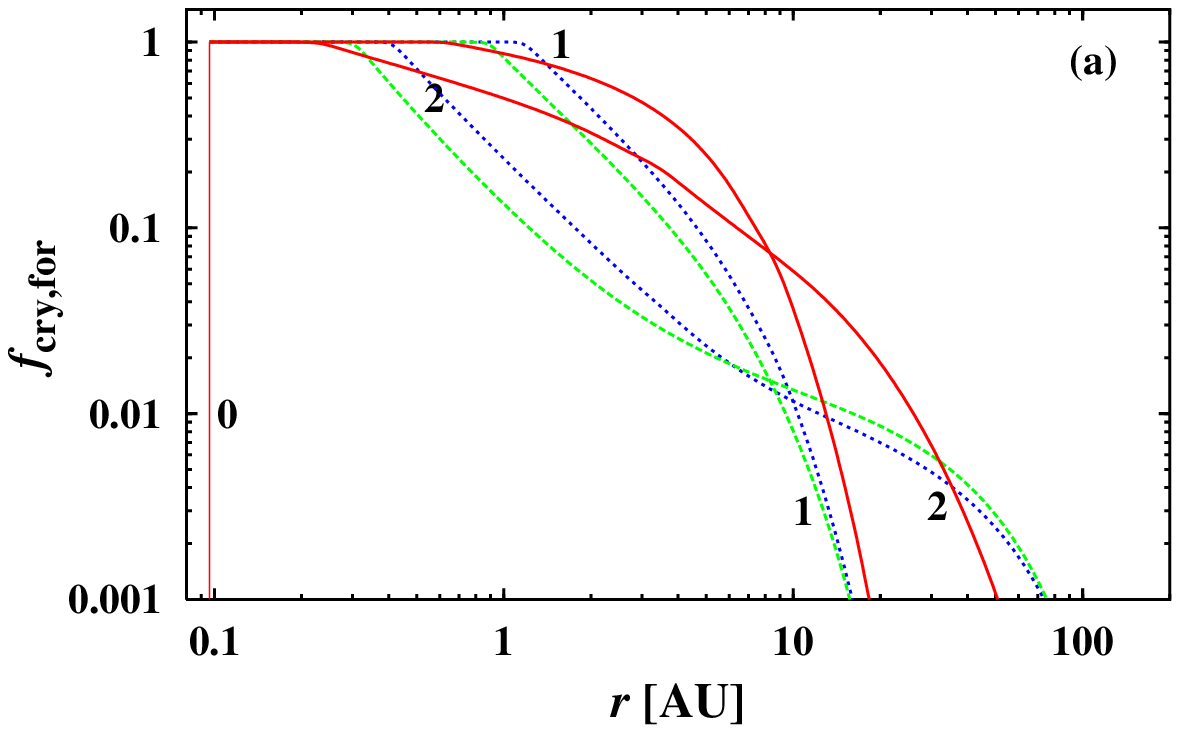}}

\resizebox{1.0\hsize}{!}{\includegraphics{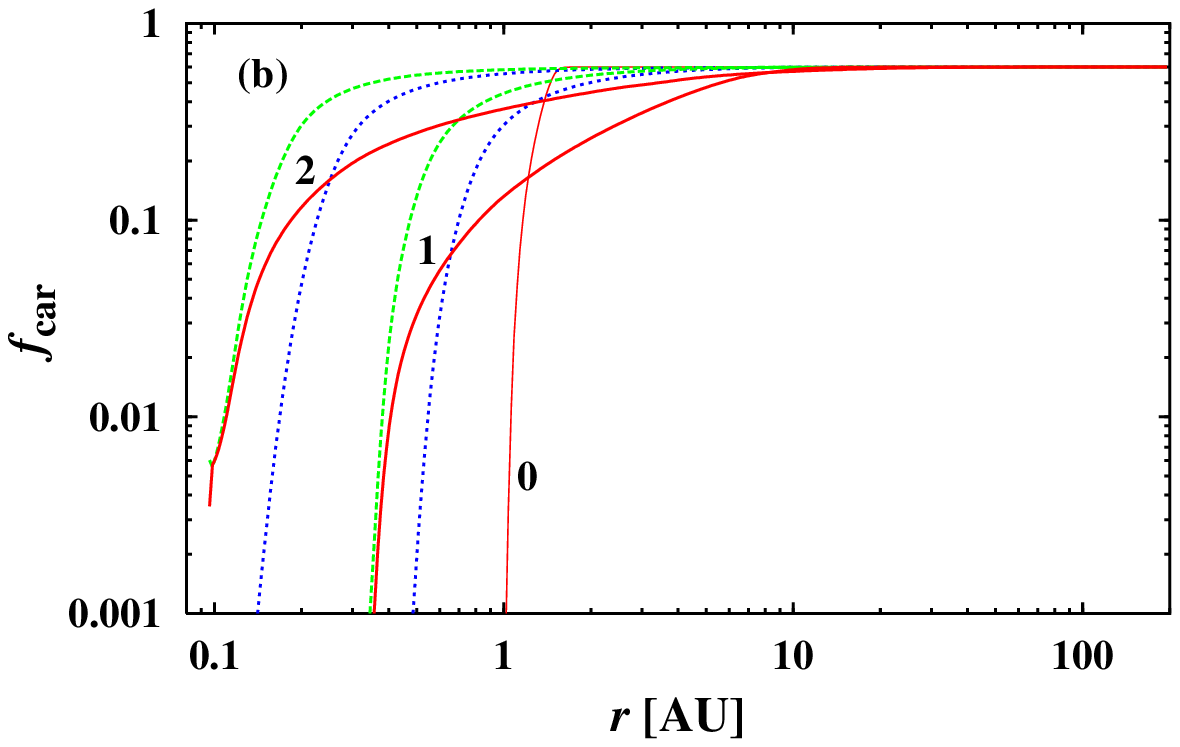}}
\caption{
Tracer transport in the $2$-dimensional models 1DE (dashed lines) and 2DM 
(solid lines) at times $t=0$ (0), $10^5$ (1) and $10^6\,{\rm yr}$ (2). 
Also shown is the result of the polar one-zone model 1DP (dashed dotted lines). 
{\bf (a)} Degree of crystallisation of forsterite $f_{\rm cry,for}$. 
{\bf (b)} Fraction of ${\rm C}$ condensed in solid carbon $f_{\rm car}$. 
}\label{figf2D}
\end{figure}

\subsubsection{Radial mixing: models 1DP and 1DE}\label{secradmix1}

We now study the impact of the meridional flow field on the efficiency of 
radial mixing processes in the disk. For this purpose in Fig.~\ref{figf2D}
the midplane radial distributions (a) of the degree of crystallisation of
forsterite $f_{\rm cry,for}$ and (b) of the fraction $f_{\rm car}$ of ${\rm C}$
condensed into solid carbon are displayed for the models 1DE (dashed lines)
and 2DM (solid lines) with the same representation as in Fig.~\ref{figf1D}. 
For a comparison the result of the polar one-zone model 1DP is plotted 
in Fig.~\ref{figf2D} with dotted lines.

\begin{figure}[t]

\resizebox{1.0\hsize}{!}{\includegraphics{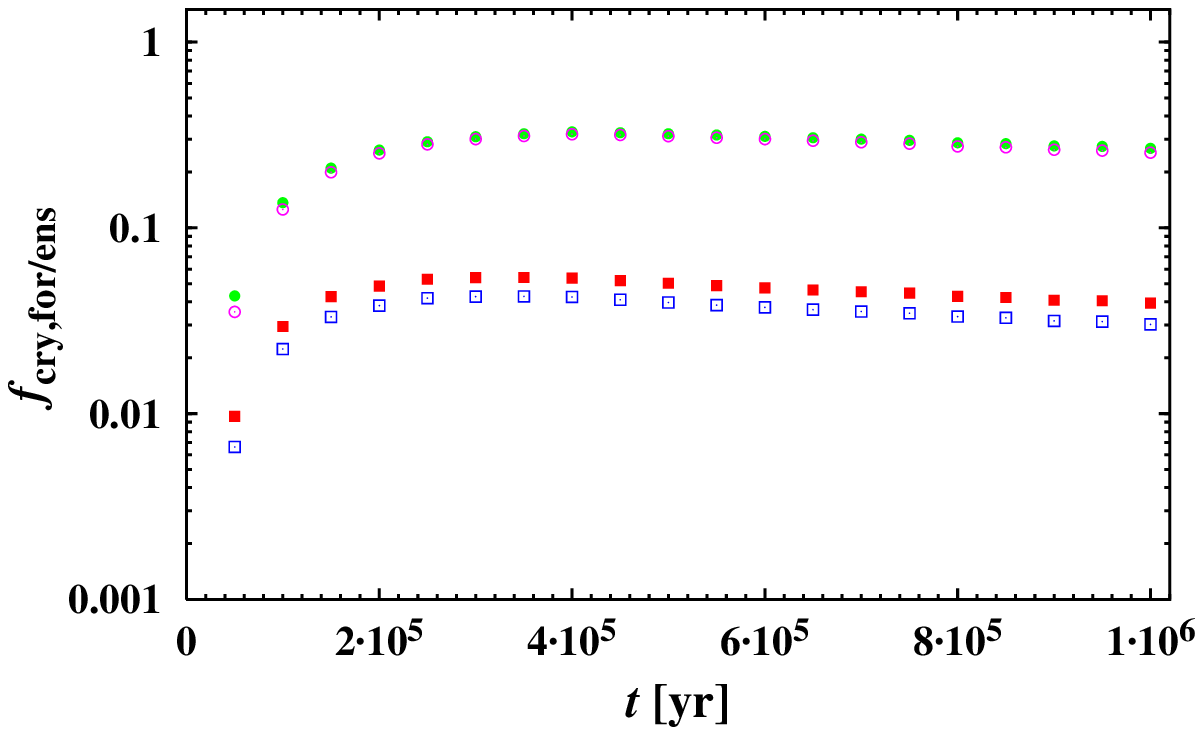}}

\caption{
Time evolution of the degrees of crystallisation of forsterite, 
$f_{\rm cry,for}$, (filled symbols) and enstatite, $f_{\rm cry,ens}$, (open 
symbols) in the models 1DE (squares) and 2DM (circles) at the position of the 
ice front. 
}\label{figfcry2Dice}
\end{figure}

We first compare the models 1DP and 1DE. In both models the tracer transport is
calculated in polar geometry with the same one-zone velocity field. Hence these
models show the influence on the tracer transport if extending it from $1$D to
$2$D. The main difference of models 1DP and 1DE can be attributed to their
different temperature structures. In the present work the models in the
$1$+$1$-dimensional approximation generally lead to lower midplane temperatures
for the chemically active parts of the disk than models in the one-zone
approximation. This is an explicit result of the calculation of the
vertical disk structure in the $1$+$1$-models, particularly of calculating the
radiative transfer in vertical direction. The midplane temperature difference
between the models 1DP and 1DE can amount to more than $100\,{\rm K}$ in the
innermost parts of the disk. As a consequence chemical reactions and dust
processing in model 1DE occur at a smaller distance $r$ than in model 1DP, i.e.,
the annealing front of silicates (Fig.~\ref{figf2D}a) as well as the carbon
combustion front (Fig.~\ref{figf2D}b) in model 1DE are shifted inward as
compared to model 1DP. 

\begin{table*}[t]
\begin{center}
\caption{
Degrees of crystallisation of forsterite and enstatite, $f_{\rm cry,for}$ and 
$f_{\rm cry,ens}$, respectively, in the model calculations 1DE and 2DM at some 
selected times $t$ and radial positions $r$. 
}\label{tabfcry}

\begin{tabular*}{\hsize}{r@{\extracolsep\fill}@{\hspace{.5cm}}|rrrr
@{\hspace{.5cm}}|rrrr}
\hline\hline
\rule[-3mm]{0mm}{8mm}& \multicolumn{4}{c|}{Model 1DE} &\multicolumn{4}{c}{Model 2DM} \\
\hline
\multicolumn{1}{c|}{$f_{\rm cry,for}$} & 
  $1\,{\rm AU}$ & $5\,{\rm AU}$ & $10\,{\rm AU}$      & $20\,{\rm AU}$      & 
  $1\,{\rm AU}$ & $5\,{\rm AU}$ & $10\,{\rm AU}$      & $20\,{\rm AU}$   
\rule[-2mm]{0mm}{6mm}   \\
\hline
\rule[0mm]{0mm}{4.5mm}
$1.0\cdot10^5\,{\rm yr}$ & 
  $0.825$       & $0.056$       & $8.0 \cdot 10^{-3}$ & $1.9 \cdot 10^{-4}$ & 
  $0.867$       & $0.248$       & $0.037$             & $4.6 \cdot 10^{-4}$ \\
$6.5 \cdot 10^5\,{\rm yr}$ & 
  $0.219$       & $0.031$       & $0.017$             & $8.9 \cdot 10^{-3}$ & 
  $0.624$       & $0.203$       & $0.075$             & $0.016$             \\
$1.0\cdot10^6\,{\rm yr}$ & 
  $0.135$       & $0.021$       & $0.013$             & $8.6 \cdot 10^{-3}$ & 
  $0.498$       & $0.134$       & $0.058$             & $0.019$             \\

\hline\hline
\multicolumn{1}{c|}{$f_{\rm cry,ens}$} & 
  $1\,{\rm AU}$ & $5\,{\rm AU}$ & $10\,{\rm AU}$      & $20\,{\rm AU}$      & 
  $1\,{\rm AU}$ & $5\,{\rm AU}$ & $10\,{\rm AU}$      & $20\,{\rm AU}$  
\rule[0mm]{0mm}{4.5mm}    \\
\hline
\rule[0mm]{0mm}{4.5mm}
$1.0\cdot10^5\,{\rm yr}$ & 
  $0.658$       & $0.043$       & $5.9 \cdot 10^{-3}$ & $1.3 \cdot 10^{-4}$ & 
  $0.841$       & $0.231$       & $0.032$             & $3.6 \cdot 10^{-4}$ \\
$6.5 \cdot 10^5\,{\rm yr}$ & 
  $0.171$       & $0.024$       & $0.014$             & $6.9 \cdot 10^{-3}$ & 
  $0.599$       & $0.196$       & $0.073$             & $0.015$             \\
$1.0\cdot10^6\,{\rm yr}$ & 
  $0.103$       & $0.016$       & $0.010$             & $6.7 \cdot 10^{-3}$ & 
  $0.471$       & $0.128$       & $0.056$             & $0.018$             \\
\hline
\end{tabular*}
\end{center}
\end{table*}
\begin{table*}[t]
\begin{center}
\caption{
Degree of condensation of carbon, $f_{\rm car}$, in the model calculations 1DE 
and 2DM at some selected times $t$ and radial positions $r$. 
}\label{tabfcar}

\begin{tabular*}{\hsize}{r@{\extracolsep\fill}@{\hspace{.5cm}}|rrrr
@{\hspace{.5cm}}|rrrr}
\hline\hline
\rule[-3mm]{0mm}{8mm}& \multicolumn{4}{c|}{Model 1DE} &\multicolumn{4}{c}{Model 2DM} \\
\hline
\multicolumn{1}{c|}{$f_{\rm car}$} & 
  $0.5\,{\rm AU}$ & $1\,{\rm AU}$ & $3\,{\rm AU}$ & $5\,{\rm AU}$ & 
  $0.5\,{\rm AU}$ & $1\,{\rm AU}$ & $3\,{\rm AU}$ & $5\,{\rm AU}$ 
\rule[-2mm]{0mm}{6mm}\\
\hline
\rule[0mm]{0mm}{4.5mm}
$1.0\cdot10^5\,{\rm yr}$ & 
  $0.132$         & $0.441$       & $0.572$       & $0.589$       & 
  $0.033$         & $0.132$       & $0.351$       & $0.473$       \\
$6.5 \cdot 10^5\,{\rm yr}$ & 
  $0.499$         & $0.566$       & $0.591$       & $0.595$       & 
  $0.188$         & $0.283$       & $0.430$       & $0.494$       \\
$1.0\cdot10^6\,{\rm yr}$ & 
  $0.546$         & $0.582$       & $0.595$       & $0.598$       & 
  $0.278$         & $0.368$       & $0.487$       & $0.571$       \\
\hline
\end{tabular*}
\end{center}
\end{table*}

Superimposed to this temperature effect a pure $2$D effect influences the radial
mixing. For evolution times $t \gtrsim 10^5\,{\rm yr}$ model 1DE shows a larger 
concentration of crystalline silicates in the outer parts of the disk than model
1DP (Fig.~\ref{figf2D}a). This can be explained by the transition from $1$- to
$2$-dimensional diffusion between both models. In model 1DP tracer particles
experience pure radial random walks owing to the $1$-dimensional treatment of
the diffusion. In contrast in model 1DE tracers are radially mixed and
additionally they are mixed in the vertical direction, i.e., the tracer particles
perform a $2$-dimensional
random walk. This leads to a more efficient radial mixing in model 1DE than in
model 1DP. In other words, concentration gradients are smoothed somewhat faster
by $2$-dimensional diffusion than by $1$-dimensional diffusion.

\subsubsection{Radial mixing: models 1DE and 2DM}\label{secradmix2}

Finally we compare the extended one-zone flow field model 1DE and the meridional
flow field model 2DM. 

Figure~\ref{figf2D} shows the remarkable impact of the meridional flow field on 
radial mixing. It clearly can be seen in the radial distribution of crystalline 
forsterite (Fig.~\ref{figf2D}a; enstatite yields a similar result). For example,
after termination of the calculations at $t = 10^6\,{\rm yr}$ the degrees of
crystallisation of forsterite, $f_{\rm cry,for}$ in model 2DM is up to $7$ times
larger than in model 1DE (at $r \approx 3\,{\rm AU}$). This result is a
consequence of the meridional flow field for which the mass flux near the
disk midplane is outward directed which makes radial mixing more efficient than
in $1$-dimensional models with pure influx. 

The region (ii) with negative $v_{r}$ at all heights (cf. Sect.~\ref{secflofie}
and Fig.~\ref{fig2Dv4}b) only slightly impedes the outward mixing of crystalline
silicates in model 2DM because it is rather narrow. In this region outward
transport is driven by diffusion only. In contrast, in region (iv) where $v_{r}
< 0$ like in region (ii), the efficiency of outward mixing in model
2DM is significantly reduced because of the large extension of this region. At
late stages of the disk evolution and for the very outer parts of the disk,
the density of annealed silicates in model 2DM even falls below that of model
1DE (at $10^6\,{\rm yr}$ for $r \gtrsim 35\,{\rm AU}$). This is a consequence
of the general outflow of matter in the one-zone velocity field of model 1DE in
the outer parts of the disk (cf. Sect.~\ref{secvelone}). The location of the
transition from accretion flow to outflow in model 1DE, $r_{x}$, is
located at  $34\,{\rm AU}$ at time $t = 10^6\,{\rm yr}$. Note that $r_{x}$
moves outward during the disk evolution due to the disk spreading. 

In Sect.~\ref{secmod1D} we mentioned that the abundance of crystalline 
silicates in the comets hardly can be explained by model calculations that are
based on a one-zone flow field and a polar disk geometry. 
Figure~\ref{figfcry2Dice} reveals that this problem does not exist for the 
meridional flow field. 
The plot shows the time evolution of the degree of crystallisation of the 
silicates (forsterite: filled symbols; enstatite: open symbols) at the ice 
front, i.e., at the radial position where water ice starts to freeze out onto 
grains (at $T \sim 150\,{\rm K}$). 
Model 1DE is represented by squares and model 2DM by circles. 
Note that the ice front slowly moves inward from $6.5\,{\rm AU}$ after 
$t = 10^5\,{\rm yr}$ to $2.5\,{\rm AU}$ at $t = 10^6\,{\rm yr}$ due to the 
accretion process. 

The degree of crystallisation of the silicates in model 2DM maintains values of
above $10\,\%$ beyond the ice front during the 
hole disk evolution. Note that the results prior to $10^5\,{\rm yr}$ are not 
physically meaningful since the disk has to evolve off from the initial model
which  is only a guess. The values of $f_{\rm cry,for}$ and $f_{\rm cry,ens}$
exceed $20\,\%$ during most of the disk evolution ($t > 150.000\,
{\rm yr}$). In contrast, in model 1DE the degree of crystallisation of the
silicates never exceeds $10\,\%$ beyond the ice front where the comets must
have been formed. Thus the abundance of crystalline silicates in many comets
of more than $10\,\%$ can be explained by outward mixing of annealed grains
from the inner disk parts into the region of comet formation. This is possible
only due to the existence of a meridional flow field. 

Note that crystalline silicates are absent in the interstellar medium (Kemper 
et al.~\cite{kem04}). Therefore it is improbable that the
crystalline silicates in the comets originate from the ISM. This points to
an origin of the crystalline silicates from the solar nebula. 

Figure~\ref{figf2D}b displays the radial distribution of condensed carbon in 
the same manner as Figure~\ref{figf2D}a. It can be seen that the narrow zone
of carbon combustion in models 1DE and 2DM for different evolutionary periods is
located at similar positions. This is a consequence of similar ${\rm OH}$
densities close to the carbon combustion front in both type of models. However,
owing to the meridional flow field, the carbon grains in model 2DM are diluted
more efficiently than in model 1DE. Hence in model 2DM a substantial amount of
the products of carbon combustion passes the ice front and is available for
incorporation into the comets. This might explain in part the large abundances of
hydrocarbons such as ${\rm CH}_4$ and ${\rm C}_2{\rm H}_2$ in comets (cf.
Paper III). 

Tables~\ref{tabfcry} and \ref{tabfcar} show values of $f_{\rm cry,for}$, 
$f_{\rm cry,ens}$, and $f_{\rm car}$ for the models 1DE and 2DM at some selected
instants $t$ and radial positions $r$. They demonstrate quantitatively the 
results discussed above. 

\begin{figure}[t]

\resizebox{1.0\hsize}{!}{\includegraphics{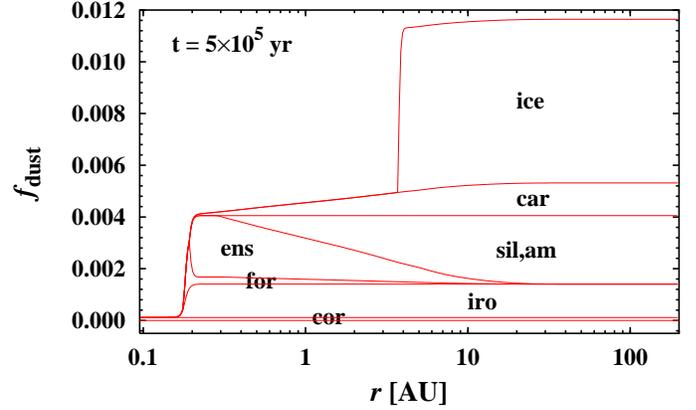}}
\caption{
Cumulative plot of the dust-to-gas mass ratios $f_{\rm dust\,:\,gas}$ of the 
individual dust species versus radial distance in model 2DM after 
$5\times 10^5\,{\rm yr}$ of disk evolution. 
The individual dust species are labeled by obvious abbreviations. 
}\label{figfcond}
\end{figure}
%

\subsubsection{Radial disk structure}\label{secradist}

The radial disk structure is substantially modified by the meridional flow 
field. The opacity of the disk matter is reduced since carbon grains
and amorphous silicate grains, that have a higher mass extinction coefficient
than crystalline silicate grains, both are less abundant
(see Fig.~\ref{figf2D}). The lower opacity results in a more efficient cooling
of the disk. As a result in the inner parts of the disk an extended radial zone
establishes where the midplane temperature in model 2DM is more than $50\,{\rm
K}$ below that of model 1DE, e.g. from $0.4$ to $2.3\,
{\rm AU}$. As the star accumulates matter from the disk this region moves
slowly radially inward with time. As a result of the lower temperature the
density is enhanced in model 2DM. As compared to model 1DE the maximum increase
of the midplane density in model  2DM amounts to $15\,\%$. 

The results show that it is essential for calculating the time dependent disk 
structure to consider both the most important opacity sources in the disk as 
well as their radial transport in the real flow field of the disk. 

\subsubsection{Radial distribution of solids}\label{secradiso}

Figure~\ref{figfcond} shows the radial distribution of solids in the disk 
for the meridional model 2DM at $5\times 10^5\,{\rm yr}$ as a cumulative plot
of the dust-to-gas mass ratios of the individual dust species. From calculations
of drag-induced drift of bodies in the solar nebula it is known that
medium-sized bodies (diameter of $\sim 10^2\,{\rm cm}$) located in the inner
disk zone rapidly spiral into the protosun on timescales of about $10^4\,
{\rm yr}$ (Weidenschilling et al. \cite{wei89}). Thus planetesimals have to be
formed quickly from these bodies. Once formed, the planetesimals remain
approximately in Keplerian orbits and mixing between different radii is not
important (Hayashi et al. \cite{hay85}). Hence plots of the kind of
Fig.~\ref{figfcond} display the composition of the planetesimals
at a given radial location if the time of their formation is known. 

Note that the headwind of the outward directed meridional flow field close to
the midplane might lengthen the timescale of drag-induced lost in the sun and potentially speeds up the formation of planetesimals.

\section{Conclusions}\label{seccon}

For a correct evaluation of the results of the model calculation we first
consider the  limitations of our present disk model. We list only the most
important issues. 

First coagulation is not considered in the present disk model. As a consequence
the dust particles remain small and are vertically mixed up to the surfaces of
the disk by turbulent diffusion, and no sedimentation of grains toward the
midplane takes place in the model. The disk structure (temperature, density
etc.) will by strongly modified by coagulation owing to a change of opacity
once this process commences. It would also be of great interest to know how
big grains are radially transported by the meridional flow field. 

Secondly a more realistic treatment of the radiative transfer in the disk is
important. This includes flaring by radiation from the protostar onto the disk
surfaces, which  is not considered in the model so far, as well as heat
transport by convection. 

At present we apply the $1$+$1$-dimensional approximation for calculating the 
disk structure while the tracer transport is computed in two dimensions. 
Hence, thirdly, exact $2$D (or even $3$D) hydrodynamics would be desirable. Such
calculations in 2D are currently underway and preliminary results are presented 
in Keller et al. (\cite{kel04}) and Tscharnuter \& Gail (\cite{Tsa07}).

Finally, in a complete model that extends the disk evolution beyond the 
first million years and comprises the chemical and mineralogical evolution of 
growing bodies, the formation of gaps in the disk due to formation of massive
planets should be included. It is to be expected that such gaps substantially
influence or even suppress radial mixing of dust and gas species across the
gaps. 

Although all these issues have to be adresses before we are able to predict,
e.g., the mineral composition of the comets or the composition of
planets, we draw the following preliminary conclusions: 
\begin{itemize}
\item 
The flow field of the disk substantially modifies the mixing of species in the 
disk. This is due to the fact that the timescales of mixing by advection is at
least of same order as that for mixing by turbulent diffusion. Hence for
tracking the spatial mixing of species, the real flow field of the disk has to be
considered in model calculations of protoplanetary disks. In the present paper
this is represented by the meridional flow field that already strongly enhances mixing compared to pure turbulent diffusion. It is to be expected that the
effect will be even more enhanced in 3D-models.
\item
The disk structure is modified by mixing processes. Species are chemically and
morphologically alterated during the disk evolution, change their opacity
properties and are mixed across the disk by diffusion and advection. This in turn
affects the radiative cooling of the disk and, hence, the disk structure. 
To account for this, at least carbon combustion and silicate annealing have to 
be incorporated in realistic models of protoplanetary disks. The transport of
these species has to be calculated self-consistently with the disk structure. 
\item
The tracer transport essentially depends on the geometry of the disk. 
Real disks flare, thus polar coordinates are more adequate for reproducing the 
disk geometry as compared to cylindrical coordinates. The flaring of the disk
causes species to be significantly depleted in the outer parts of the disk as
compared to the case of a flat geometry. Note that cylindrical coordinates have
mostly been used in previous disk models that include tracer transport. 
\item
The depletion of tracers in models with polar disk geometry in the outer disk 
regions, particularly for annealed silicates, is more than compensated by
the efficient outward mixing of these species via the meridional flow field. 
Therefore the presence of a fraction of crystalline silicates in some comets
of more than $10\,\%$ can easily be explained by the present model calculations.
In contrast, the concentration of crystalline silicates in the region of comet 
formation is more than an order of magnitude lower in models based on the
one-zone flow field. This confirms also the results of the $2$-dimensional model
calculations of Keller \& Gail~(\cite{kel04}). The large abundance of methane
and acetylene in the comets also could, at least in part, be due to the outward
transport of exhaustion gases by the meridional flow field. 
\end{itemize}

The results of the present model calculations have to be confirmed by future 
hydrodynamic disk models. Fully 2-D radiation hydrodynamic models are presently in
preparation, some preliminary results of which are already published in 
Tscharnuter \& Gail (\cite{Tsa07}), but such calculations presently cannot compete 
with the long time basis and spatial resolution of model calculations that can be obtained with 1+1-dim models like the present one.

As a next step, this model will be coupled with a detailed modeling of the chemistry in the disk. 

\begin{acknowledgements}
This work has been supported by the Deutsche Forschungsgemeinschaft (DFG), 
Sonderforschungsbereich 359 `Reactive Flows, Diffusion and Transport', 
and by a PhD grant of the Landesgraduiertenf\"orderung Baden W\"urttemberg. 
\end{acknowledgements}



\begin{thebibliography}{}

\bibitem[2007]{Ale07} Alexander, C.~M.~O., Boss, A.~P., Keller, L.~P.,
  Nuth,J.~A., \& Weinberger, A.\ 2007, Protostars and Planets V, 801 
\bibitem[2002]{boc02} Bockel\'ee-Morvan, D., Gautier, D., Hersant, F., et al. 
  2002, \aap, 384, 1107
\bibitem[2004]{bos04} Boss, A.~P.\ 2004, \apj, 616, 1265 
\bibitem[2007]{bos07} Boss, A.~P.\ 2007, \apj, 660, 1707
\bibitem[2008]{bos08} Boss, A.~P.\ 2008, ArXiv e-prints, 801, arXiv:0801.1622
\bibitem[2001]{Bou01} Bouwman, J., Meeus, G., de Koter, A., Hony,
  S., Dominik, C., \& Waters, L.~B.~F.~M.\ 2001, \aap, 375, 950 
\bibitem[1997]{cro97} Crovisier, J., Leech, K., Bockel\'{e}e-Morvan, D., et al. 
  1997, Science, 275, 1904
\bibitem[1998]{cyr98} Cyr, K.\,E., Sears, W.\,D., \& Lunine, J.\,I. 1998, 
  \icarus, 135, 537
\bibitem[1999]{dro99} Drouart, A., Dubrulle, B., Gautier, D., et al. 1999, 
  \icarus, 140, 129
\bibitem[2001]{gai01} Gail, H.-P. 2001, \aap, 378, 192 (Paper I)
\bibitem[2002]{gai02} Gail, H.-P. 2002, \aap, 390, 253 (Paper III)
\bibitem[2004]{gai04} Gail, H.-P. 2004, \aap, 413, 571 (Paper IV)
\bibitem[1985]{hay85} Hayashi, C., Nakazawa, K., \& Nakagawa, Y. 1985, in
  Formation of the Solar System, ed.  D.~C. Black, \& M.~A. Shapley (Tucson:
University of Arizona Press), 1100
\bibitem[1997]{han97} Hanner, M.\,S., Gehrz, R.\,D., Harker, D.\,E., et al. 
  1997, Earth, Moon and Planets, 79, 247
\bibitem[1994]{han94} Hanner, M.\,S., Lynch, D.\,K., \& Russel, R.\,W. 1994, 
  \apj, 425, 274
\bibitem[1964]{hir64} Hirschfelder, J.\,O., Curtiss, C.\,F., \& Bird, R.B. 
  1964, Molecular Theory of Gases and Liquids, Wiley, New York
\bibitem[2005]{joh05} Johansen, A., \& Klahr, H.\ 2005, \apj, 634, 1353
\bibitem[2006]{joh06} Johansen, A., Klahr, H., \& Mee, A.~J.\ 2006, \mnras, 370,
  L71
\bibitem[2003]{kel03} Keller, Ch. 2003, Ph.D. thesis,
  Ruprecht-Karls-Universit\"at, Heidelberg, Germany
\bibitem[2004]{kel04} Keller, Ch., \& Gail, H.-P. 2004, \aap, 415, 1177 (Paper 
  VI)
\bibitem[2005]{Kel05} Keller, L.~P., \& Messenger, S.\ 2005, Chondrites and the
  Protoplanetary Disk, 341, 657 
\bibitem[2004]{kem04} Kemper, F., Vriend, W.\,J., \& Tielens, A.\,G.\,G.\,M. 
  2004, \apj, 609, 826
\bibitem[2004]{kla04} Klahr, H.\,H. 2004, \apj, 606, 1070
\bibitem[2003]{kla03} Klahr, H.\,H., \& Bodenheimer, P. 2003, \apj, 582, 869
\bibitem[1992]{kle92} Kley, W., \& Lin, D.\,N.\,C. 1992, \apj, 397, 600
\bibitem[2000]{klu00} Klu\'zniak, W., \& Kita, D. 2000, preprint 
  {\tt [astro-ph/0006266]}
\bibitem[1985]{lin85} Lin, D.\,N.\,C., \& Papaloizou, J. 1985, in Protostars \& 
  Planets II, ed. D.\,C. Black, \& M.\,S. Matthews (Tucson: University of
  Arizona Press), 981
\bibitem[2001]{Meu01} Meeus, G., Waters, L.~B.~F.~M., Bouwman, J., van den Ancker,
  M.~E., Waelkens, C., \& Malfait, K.\ 2001, \aap, 365, 476 
\bibitem[1978]{mih78} Mihalas, D. 1978, Stellar Atmospheres (San Francisco:
 W.\,H. Freeman \& Co.) 
\bibitem[1978]{pac78} Paczy\'nski, B. 1978, \actaa, 28, 91
\bibitem[1992]{pre92} Press, W.\,H., Teukolsky, S.\,A., Vetterling, W.\,T., et 
  al. 1992, Numerical Recipes in FORTRAN, Second   Edition (Cambridge: Cambridge
  University Press)
\bibitem[2007]{pav07} Pavlyuchenkov, Y., \& Dullemond, C.~P.\ 2007, \aap, 471, 833 
\bibitem[1981]{pri81} Pringle, J.\,E. 1981, \araa, 19, 137
\bibitem[2002]{reg02} Regev, O., \& Gitelman, L. 2002, \aap, 396, 623
\bibitem[2004]{ric04} Richard, D., \& Davis, S.\,S. 2004, \aap, 416, 825
\bibitem[1994]{roz94} R\'o\.zyczka, M., Bodenheimer, P., \& Bell, K.\,R. 1994, 
  \apj, 423, 736
\bibitem[1986]{rud86} Ruden, S.\,P., \& Lin, D.\,N.\,C. 1986, \apj, 308, 883
\bibitem[1988]{sie88} Siemiginowska, A. 1988, \actaa, 38, 21
\bibitem[1988]{ste88} Stevenson, D.\,J., \& Lunine, J.\,I. 1988, \icarus, 75, 146
\bibitem[1988]{Swa88} Swamy, K.~K.~S., Sandford, S.~A., Allamandola, L.~J.,
  Witteborn, F.~C., \& Bregman, J.~D.\ 1988, Icarus, 75, 351 
\bibitem[2002]{tak02} Takeuchi, T., \& Lin, D.\,N.\,C. 2002, \apj, 581, 1344
\bibitem[2007]{Tsa07} Tscharnuter, W.\,M., \& Gail. H.-P. 2007, \aap, 463, 369
\bibitem[2006]{tur06} Turner, N.~J., Willacy, K., Bryden, G., \& Yorke, H.~W.\ 
  2006, \apj, 639, 1218 
\bibitem[1984]{urp84} Urpin, V.\,A. 1984, \sovast, 28, 50
\bibitem[2004]{vBo04} van Boekel, R., et al.\ 2004, \nat, 432, 479 
\bibitem[2005]{vBo05} van Boekel, R., Min, M., Waters, L.~B.~F.~M., de Koter, A.,
  Dominik, C., van den Ancker, M.~E., \& Bouwman, J.\ 2005, \aap, 437, 189
\bibitem[2003]{weh03b} Wehrstedt, M. 2003, Ph.D. thesis, 
  Ruprecht-Karls-Universit\"at, Heidelberg, Germany
\bibitem[2002]{weh02} Wehrstedt, M., \& Gail, H.-P. 2002, \aap, 385, 181 (Paper 
  II)
\bibitem[2003]{weh03} Wehrstedt, M., \& Gail, H.-P. 2003, \aap, 410, 917 (Paper 
  V)
\bibitem[1989]{wei89} Weidenschilling, S.~J., Donn, B., \& Meakin, P. 1989, in
  The Formation and Evolution of Planetary Systems, ed. H.~A. Weaver, \& L.
  Danly (Cambridge: Cambridge University Press), 131
\bibitem[2005]{woo05} Wooden, D.~H., Harker, D.~E., \& Brearley, A.~J.\ 2005,
Chondrites and the Protoplanetary Disk, 341, 774 
\bibitem[2007]{woo07} Wooden, D., Desch, S., Harker, D., Gail, H.-P., \& Keller,
  L.\ 2007, Protostars and Planets V, 815
\bibitem[1999]{yan99} Yanamandra-Fisher, P.\,A., \& Hanner, M.\,S. 1999, 
  \icarus, 138, 107

\end{thebibliography}
\end{document}